\pdfoutput=1
\documentclass[11pt,twoside,a4paper,cmspaper,final,collab]{cms-tdr}
\def\svnVersion{3e99516}\def\svnDate{2022/03/28}\def\cmsCernNoTag{CERN-EP-2021-115}\def\cmsCernDate{\today}\def\cmsMessage{Published in the Journal of High Energy Physics as \href{http://dx.doi.org/10.1007/JHEP05(2022)093}{\doi{10.1007/JHEP05(2022)093}.}}
\begin{document}\cmsNoteHeader{B2G-19-004}

\newcommand{\tZ}{\ensuremath{\PQt\PZ}\xspace}
\newcommand{\wjets}{\PW{+}\text{jets}\xspace}
\newcommand{\zjets}{\PZ{+}\text{jets}\xspace}
\newcommand{\ttjets}{\ensuremath{\PQt\PAQt}+\text{jets}\xspace}
\newcommand{\MT}{\ensuremath{M_{\PQT}}\xspace}

\cmsNoteHeader{B2G-19-004} 

\title{Search for single production of a vector-like \texorpdfstring{\PQT}{T} quark decaying to a top quark and a \texorpdfstring{\PZ}{Z} boson in the final state with jets and missing transverse momentum at \texorpdfstring{$\sqrt{s} = 13\TeV$}{sqrt(s) = 13 TeV}}

\date{\today}

\newlength\cmsTabSkip\setlength{\cmsTabSkip}{1ex}

\abstract{
A search is presented for single production of a vector-like \PQT quark with charge~$2/3$\,$e$, in the decay channel featuring a top quark and a \PZ boson, with the top quark decaying hadronically and the \PZ boson decaying to neutrinos. The search uses data collected by the CMS experiment in proton-proton collisions at a center-of-mass energy of 13\TeV, corresponding to an integrated luminosity of 137\fbinv recorded at the CERN LHC in 2016--2018. The search is sensitive to a \PQT quark mass between 0.6 and 1.8\TeV with decay widths ranging from negligibly small up to 30\% of the \PQT quark mass. Reconstruction strategies for the top quark are based on the degree of Lorentz boosting of its final state. At 95\% confidence level, the upper limit on the product of the cross section and branching fraction for a \PQT quark of small decay width varies between 15 and 602\unit{fb}, depending on its mass. For a \PQT quark with decay widths between 10 and 30\% of its mass, this upper limit ranges between 16 and 836\unit{fb}. For most of the studied range, the results provide the best limits to date. This is the first search for single \PQT quark production based on the full Run 2 data set of the LHC.
}

\hypersetup{%
pdfauthor={CMS Collaboration},%
pdftitle={Search for single production of a vector-like T quark decaying to a top quark and a Z boson in the final state with jets plus missing transverse momentum at sqrt(s) = 13 TeV},%
pdfsubject={CMS},%
pdfkeywords={CMS, vector-like quarks}}

\maketitle 

\section{Introduction}\label{sec:intro}
New heavy color-triplet spin-$1/2$ fermions with nonchiral couplings, referred to as vector-like quarks (VLQs), are predicted in many extensions of the standard model (SM)~\cite{Aguilar-Saavedra:2013qpa, Okada:2012gy, AguilarSaavedra:2009es, DeSimone:2012fs, Matsedonskyi:2014mna, Buchkremer:2013bha} in order to resolve theoretical issues, such as the hierarchy problem. While the masses of the SM chiral quarks arise from Yukawa couplings to the Higgs field, for VLQs non-Yukawa coupling terms in the Lagrangian are allowed.
The existence of VLQs is not yet excluded by precision SM measurements, unlike the case of chiral quarks from a fourth generation~\cite{Eberhardt:2012sb}.
Pair production of VLQs occurs via strong interaction processes while single production, which dominates for VLQ masses above $\approx$1\TeV owing to the larger phase space, occurs via electroweak interaction processes.

\begin{figure}[!bh]
\centering
\includegraphics[scale=0.40]{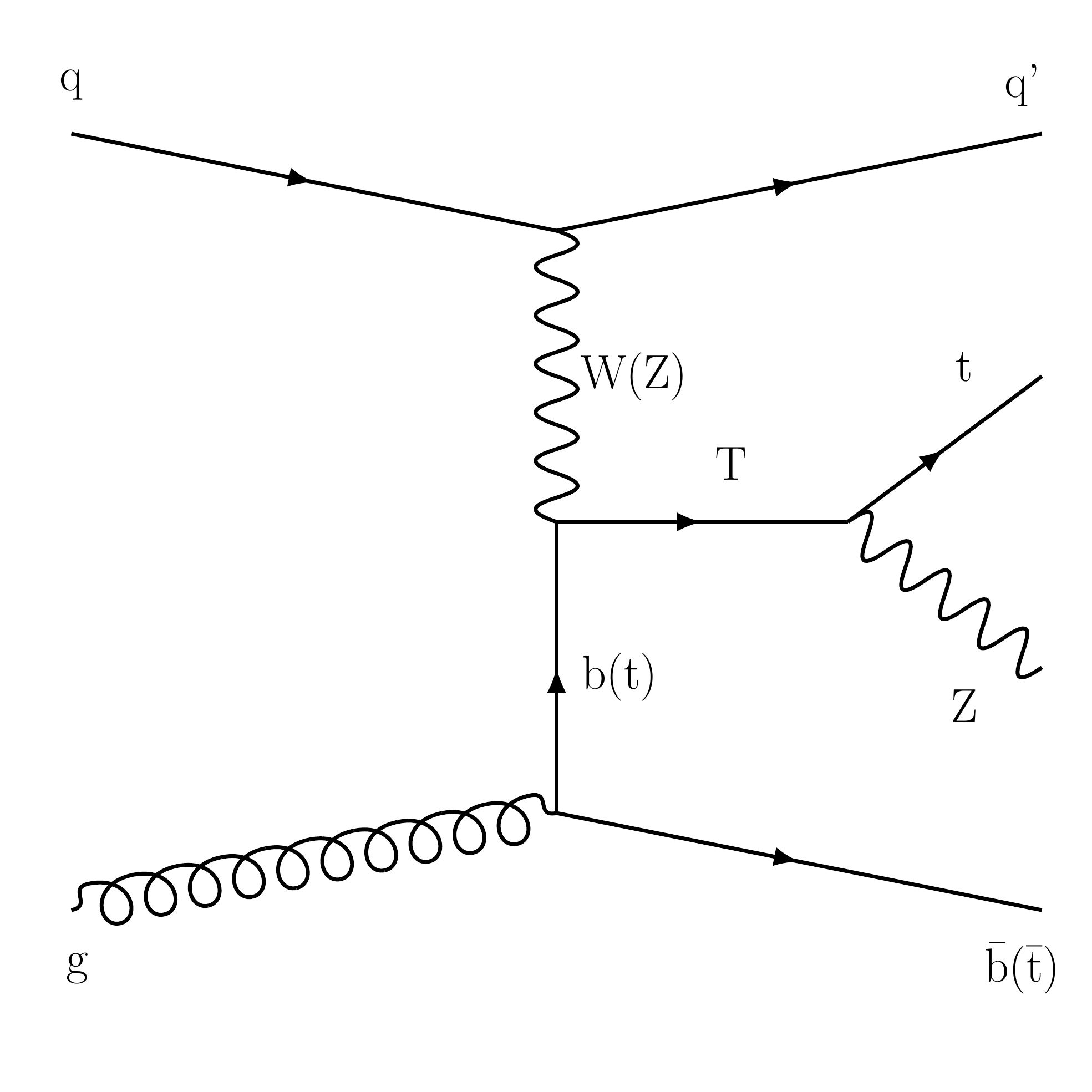}
\caption{A representative leading-order Feynman diagram for the production of a single vector-like quark \PQT decaying into a \PZ boson and a top quark.}\label{fig:feynman}
\end{figure}

This paper presents a search for the single production of a vector-like \PQT quark with charge~$2/3$\,$e$, where the \PQT quark decays to a top quark (\PQt) and a \PZ boson.
The analysis is performed in proton-proton ($\Pp\Pp$) collision data at $\sqrt{s}=13\TeV$ collected with the CMS detector in Run 2 of the CERN LHC in 2016--2018.
An example of a leading-order (LO) Feynman diagram for single \PQT quark production and decay is shown in Fig.~\ref{fig:feynman}.

This analysis targets $\PQt\PZ$ final states where the top quark decays hadronically via $\PQt \to \PW\PQb \to \PQq^{\prime}\PAQq\PQb$ and the \PZ boson decays to neutrinos. Neutrinos are not detected in the experimental apparatus, therefore a full reconstruction of the \PQT quark four-momentum cannot be performed, and signal events are characterized by a large transverse momentum (\pt) imbalance.

In addition to the $\PQt\PZ$ decay channel, a \PQT quark can also decay to a bottom quark (\PQb) and a \PW boson, or a top quark and a Higgs boson ($\PH$), with branching fractions that are a function of the model-dependent couplings with SM third-generation quarks. 
For a singlet \PQT quark model~\cite{AguilarSaavedra:2009es} the branching fractions for the $\PQb\PW$, $\PQt\PZ$, and $\PQt\PH$ decay channels are approximately 0.5, 0.25, and 0.25, respectively.

Searches for pair production of \PQT quarks at 13\TeV, targeting a variety of possible final states from the above decay channels, have been performed by the ATLAS~\cite{Aaboud:2017qpr, Aaboud:2017zfn, Aaboud:2018saj, Aaboud:2018wxv, Aaboud:2018xuw, Aaboud:2018pii} and CMS~\cite{Sirunyan:2017usq, Sirunyan:2017pks, Sirunyan:2018omb, Sirunyan:2018qau, Sirunyan:2019sza} Collaborations. In particular, the paper~\cite{Aaboud:2018pii} presenting the combination of the searches at 13\TeV by the ATLAS Collaboration has set a lower limit of 1.3\TeV at 95\% confidence level on the mass of a \PQT quark in the singlet \PQT quark model~\cite{AguilarSaavedra:2009es}.

A search for single production of \PQT quarks exploiting the same final state as in this analysis, and based on the 2015--2016 data set only, has been reported by the ATLAS Collaboration~\cite{Aaboud:2018zpr}. Searches in the $\PQt\PZ$ channel using final states where the Z boson decays to a pair of muons or electrons have been published by both the ATLAS and CMS Collaborations~\cite{Aaboud:2018saj, Sirunyan:2017ynj}. A search exploiting fully hadronic final states, targeting $\PQt\PZ$ and $\PQt\PH$ channels, has been reported in a recent paper~\cite{Sirunyan:2019xeh} by the CMS Collaboration. Also the $\PQb\PW$ channel has been investigated at 13\TeV in final states where the \PW boson decays leptonically~\cite{Aaboud:2018ifs, Sirunyan:2017tfc}. No deviations have been observed in any of these searches with respect to the expected SM background. All the above searches are based on data sets corresponding to a maximum integrated luminosity of about 36\fbinv, which is almost a factor of four less than the 137\fbinv used for the analysis described in this paper.

As shown in Fig.~\ref{fig:feynman}, final states of signal events are characterized by a top or bottom quark produced in association with the \PQT quark as well as a quark that tends to be emitted at a low angle with respect to the beam axis, in the forward region of the detector. For the interpretation of our search, we assume a singlet \PQT quark model~\cite{Matsedonskyi:2014mna} in which the production process is the associated production with a bottom quark. This process has a much larger cross section, the production in association with a top quark being suppressed owing to the higher top quark mass~\cite{Matsedonskyi:2014mna}.
However, no specific requirement on the associated top or bottom quark is considered in the analysis selection and strategy, and by design the results of this search may also be interpreted in terms of other models with a single top quark and missing transverse momentum in the final state.

The considered mass values for the \PQT quark range from 0.6 to 1.8\TeV.
Multiple hypotheses for the \PQT quark width ($\Gamma$) are probed for the first time in this channel, with values varying from 1 to 30\% of the \PQT quark mass, by employing simulated signal samples generated using different width hypotheses.
A novel and distinctive feature of this search is that the analysis exploits the fact that the \PQT quark's final state appears differently in the detector, depending on its mass.  Accordingly, different reconstruction strategies for the top quark are employed for different mass ranges of the \PQT quark. For lower mass values the quarks from the top quark decay tend to be reconstructed as individual, narrow-cone jets.
As the \PQT quark mass increases, the decay products of the boosted top quark become highly collimated, producing jets
that overlap either completely or partially.
Therefore, for the case of large \PQT quark masses, large-radius jets are studied
with the aid of substructure techniques in order to identify jets originating from the top quark or the \PW boson, 
enhancing the analysis sensitivity. 

Tabulated results are provided in the HEPData record for this analysis~\cite{hepdata}.

\section{The CMS detector, data, and simulation}\label{sec:detector}
This analysis is based on $\Pp\Pp$ collision data at $\sqrt{s}=13\TeV$ collected with the CMS detector in Run 2 of the LHC during the years 2016--2018, and corresponding to integrated luminosities of 36.3, 40.7, and 59.8\fbinv, for the 2016, 2017, and 2018 data sets, respectively.
The central feature of the CMS apparatus is a superconducting solenoid of 6\unit{m} internal diameter, providing a magnetic field of 3.8\unit{T}. Within the solenoid volume are a silicon pixel and strip tracker, a lead tungstate crystal electromagnetic calorimeter (ECAL), and a brass and scintillator hadron calorimeter (HCAL), each composed of a barrel and two endcap sections. Forward calorimeters, made of steel and quartz-fibres, extend the pseudorapidity ($\eta$) coverage provided by the barrel and endcap detectors. Muons are detected in gas-ionization chambers embedded in the steel flux-return yoke outside the solenoid.
A more detailed description of the CMS detector, together with a definition of the coordinate system used and the relevant kinematic variables, can be found in Ref.~\cite{Chatrchyan:2008zzk}. 

Events of interest are selected using a two-tiered trigger system. The first level (L1), composed of custom hardware processors, uses information from the calorimeters and muon detectors to select events at a rate of around 100\unit{kHz} within a fixed latency of about 4\mus~\cite{Sirunyan:2020zal}. The second level, known as the high-level trigger (HLT), consists of a farm of processors running a version of the full event reconstruction software optimized for fast processing, and reduces the event rate to around 1\unit{kHz} before data storage~\cite{Khachatryan:2016bia}. 
Signal events of this analysis, because of the presence of a boosted Z boson decaying to a pair of neutrinos,
are selected online by triggers which require a large transverse momentum imbalance.

Simulated background samples are generated at LO with \MGvATNLO \cite{Alwall:2014hca} for \wjets, \zjets, and for events with jets arising exclusively from quantum chromodynamics (QCD) interactions (multijet events). The next-to-LO (NLO) generator \POWHEG 2.0 \cite{Nason:2004rx, Frixione:2007vw, Alioli:2010xd} is used to simulate top quark-antiquark pairs produced in association with jets (\ttjets)~\cite{Frixione:2007nw} and single top quark~\cite{fourfiveflavorschemes} events. The \MGvATNLO version is 2.2.2 (2.4.2) for simulated events used with 2016 (2017 and 2018) data. 
Diboson events with inclusive final states are generated at LO using \PYTHIA 8.2 \cite{Sjostrand:2014zea}. Diboson events with exclusive final states, generated at NLO with \MGvATNLO and \POWHEG 2.0, are employed in specific signal regions defined in the analysis to improve the statistical precision of the background estimates.
Finally, \MGvATNLO at NLO is used for \ttbar{+}V and tZq samples, which are minor backgrounds.
The \POWHEG 2.0 and \MGvATNLO generators are interfaced with \PYTHIA 8.2 with the following versions and underlying event tunes: version 8.226 and tune CUETP8M1 \cite{Khachatryan:2015pea} for the simulated events used with 2016 data, and version 8.230 and tune CP5 \cite{Sirunyan:2019dfx} for those used with 2017 and 2018 data. The event tune CUETP8M2T4 \cite{CMS-PAS-TOP-16-021} is employed for the simulated \ttjets events used with 2016 data, and the \PYTHIA version 8.240 is adopted for simulated \ttjets events used for 2017 and 2018 data. 
The background processes are initially normalized to their theoretical cross sections, using the highest order available.
The \wjets and \zjets events are further normalized to account for NLO in electroweak and next-to-NLO (NNLO) in QCD corrections to the cross sections from Ref.~\cite{Lindert:DYCorrections}.
The cross section for the \ttbar background is computed at NNLO in perturbative QCD using a soft-gluon resummation at next-to-next-to-leading logarithmic accuracy with the Top++ 2.0 program \cite{Czakon:2011xx}. 
For single top quark production, NLO calculations are performed with \textsc{HATHOR} 2.1~\cite{hator}.

Signal event samples are generated in the 4-flavor scheme at LO using \MGvATNLO interfaced to \PYTHIA 8.2. A singlet \PQT quark is assumed that is produced in association with a bottom quark with suppressed right-handed coupling to SM particles~\cite{Cacciapaglia:2011fx}. The contribution of the associated production with a top quark is within the theory uncertainty of the dominant production with a bottom quark and is considered negligible.
The generated \PQT quark mass in the samples ranges from 0.6 up to 1.8\TeV, and several hypotheses for the \PQT quark width are considered. Narrow-width samples are generated in mass steps of 0.1\TeV and with $\Gamma$ set to 10\GeV, which is much smaller than the experimental resolution. Large-width samples are generated in mass steps of 0.2\TeV with $\Gamma$ set to 10, 20, and 30\% of the resonance mass $m_{\PQT}$. Again, the version \MGvATNLO~2.2.2 (2.4.2) is used for 2016 (2017 and 2018) simulated events. The choice of the interval and granularity for the large-width samples, both in $m_{\PQT}$ and $\Gamma$, takes into account the sensitivity of the discriminating observables to these variables.
All possible decays of the top quark and $\PZ$ boson have been considered for the signal process. 
The total cross section for a singly produced \PQT quark that decays to $\PQt\PZ$ is computed using the interpretation framework described in Ref.~\cite{Sirunyan:2018BtoHb}.
The cross section values for the $\Gamma/m_\PQT$ hypotheses of 1, 10, 20, and 30\% are obtained from computations performed at LO QCD with the full width calculation, based on Ref.~\cite{PhysRevD.98.015029}.
The cross section values for the single production of a vector-like singlet \PQT quark with $\Gamma/m_\PQT$ of 1, 10, 20, and 30\% are reported in Table~\ref{tab:CrossSec}. 
An additional benchmark is considered for a value of $\Gamma/m_\PQT$ of 5\%, and the same narrow width signal sample is used in this case as for the hypothesis with 1\% for $\Gamma/m_\PQT$. The cross section values for the 5\% mass points are extrapolated from the values corresponding to the 1 and 10\% hypotheses using the procedure described in Ref.~\cite{Sirunyan:2019xeh}.

\begin{table}[!h]
\topcaption{Theoretical cross sections, in fb, for single production of a vector-like singlet \PQT quark in association with a bottom quark, with the \PQT quark decaying to $\PQt\PZ$, for mass hypotheses between 0.6 and 1.8\TeV in steps of 0.1\TeV, and for resonance widths that are 1, 10, 20, and 30\% of its mass. The framework for the computation is described in Refs.~\cite{Sirunyan:2018BtoHb,PhysRevD.98.015029}. The uncertainties, given in parentheses, are obtained by halving and doubling the values of QCD renormalization and factorization scales.}\label{tab:CrossSec}
\centering
\begin{tabular}{ccccc}
\multicolumn{5}{c}{$\sigma$($\Pp\Pp\to\PQT\PQb\PQq\to\PQt\PZ\PQb\PQq$) [fb]}\\[\cmsTabSkip]
$m_\PQT$ [\TeVns{}] & $\Gamma/m_\PQT = 1\%$ & $\Gamma/m_\PQT = 10\%$ & $\Gamma/m_\PQT = 20\%$ & $\Gamma/m_\PQT = 30\%$ \\
\hline
0.6 & $175\,(34)$    & $1750\,(340)$ & $3500\,(690)$    & $5200\,(1000)$\\
0.7 & $78\,(16)$     & $780\,(160)$  & $1520\,(310)$    & $2240\,(710)$\\
0.8 & $41.5\,(8.6)$  & $409\,(85)$   & $790\,(170)$     & $1150\,(250)$\\
0.9 & $23.3\,(5.0)$  & $228\,(49)$   & $437\,(94)$      & $630\,(140)$\\
1.0 & $13.6\,(3.0)$  & $132\,(29)$   & $251\,(55)$      & $360\,(80)$\\
1.1 & $8.2\,(1.8)$   & $79\,(18)$    & $150\,(34)$      & $214\,(48)$\\
1.2 & $5.1\,(1.2)$   & $49\,(11)$    & $93\,(21)$       & $131\,(30)$\\
1.3 & $3.25\,(0.75)$ & $31.2\,(7.3)$ & $59\,(14)$       & $82\,(19)$\\
1.4 & $2.12\,(0.50)$ & $20.3\,(4.8)$ & $37.9\,(9.0)$    & $54\,(13)$\\
1.5 & $1.41\,(0.34)$ & $13.4\,(3.2)$ & $25.0\,(6.0)$    & $35.2\,(8.4)$\\
1.6 & $0.94\,(0.23)$ & $9.0\,(2.2)$  & $16.8\,(4.1)$    & $23.5\,(5.7)$\\
1.7 & $0.64\,(0.16)$ & $6.2\,(1.5)$  & $11.5\,(2.8)$    & $16.0\,(3.9)$\\
1.8 & $0.45\,(0.11)$ & $4.2\,(1.1)$ & $7.9\,(2.0)$     & $11.0\,(2.8)$\\
\end{tabular}
\end{table} 

The simulated samples are generated using the LO NNPDF~3.0~\cite{Ball:2014uwa} parton distribution function (PDF) sets for 2016, except for \ttbar and single top quark events, which adopt NLO PDF sets in this case, and using the NNLO NNPDF~3.1~\cite{Ball:2017nwa} PDF sets for 2017 and 2018.
The generated events are passed through a simulation of the CMS detector based on \GEANTfour~\cite{Agostinelli:2002hh, Allison:2006ve}.
Additional interactions in the same or adjacent bunch crossings (pileup) are included in the simulation, with their frequency distribution adjusted to match that observed in data.

\section{Event reconstruction and selection}\label{sec:reconstruction}
The candidate vertex with the largest value of summed physics-object $\pt^2$ is taken to be the primary $\Pp\Pp$ interaction vertex. The physics objects are the jets, clustered using the jet finding algorithm~\cite{Cacciari:2008gp,Cacciari:2011ma} with the tracks assigned to candidate vertices as inputs, and the associated missing transverse momentum, taken as the negative vector sum of the \pt of those jets.
The primary vertex must be within 24\unit{cm} of the nominal interaction point along the beam axis and within 
2\unit{cm} in the transverse plane.

Particle candidates in the event are reconstructed using the particle-flow (PF) algorithm~\cite{Sirunyan:2017ulk}, 
which performs a global event reconstruction by combining the information from the various elements of the CMS detector, 
and which provides identification of muons, electrons, photons, and charged and neutral hadrons.
Jets are clustered from the PF candidates using the anti-\kt algorithm~\cite{Cacciari:2008gp} with 
distance parameters of 0.4 (AK4 jets) and 0.8 (AK8 jets). 
An algorithm~\cite{Cacciari:2007fd} is applied to the AK4 jets to remove
charged hadrons not originating from the primary vertex, while for the AK8 jets the pileup per 
particle identification algorithm~\cite{PUPPI} is employed, which assigns
a weight to charged and neutral PF candidates, according to the likelihood
that the particle originates from a pileup interaction.
The weight is then used to rescale the particle four-momentum.
Correction factors as a function of the \pt, $\eta$, energy density, and the area of the jet 
are applied to calibrate the jet energy scale.
The jet energy resolution for simulated jets is adjusted to reproduce the resolution in data~\cite{Khachatryan:2016kdb}.

The missing transverse momentum vector \ptvecmiss is defined as the negative vector \pt sum
of all the PF candidates in an event, and its magnitude is denoted as \ptmiss~\cite{Sirunyan:2019kia}.
The missing hadronic activity \htvecmiss is defined as the negative vector sum of the AK4 jets \pt,
and its magnitude is denoted as \mht.
The \ptvecmiss and \htvecmiss are modified to account for corrections to the energy scale of the reconstructed jets in the event.
The missing transverse momentum and the missing hadronic activity are computed also at online level, using algorithms similar to the offline versions. Events are selected at the HLT if they have \ptmiss and \mht greater than 120\GeV.

At offline level, the AK4 jets selected for the analysis must have $\pt>30\GeV$, while the AK8 jets must have $\pt>200\GeV$.
Central AK4 or AK8 jets must have $\abs{\eta} < 2.4$. Central AK4 jets are used as input to the \PQb tagging algorithm
to identify jets originating from the hadronization of a bottom quark.
Forward jets are defined as AK4 jets in the region $2.4 \le \abs{\eta} \le 4.0$.
Jets with larger values of $\abs{\eta}$ are not included in the count of forward jets, in order to reduce the contribution
from pileup interactions.

The mass of an AK8 jet is measured after applying jet grooming techniques, 
in which the constituents of the AK8 jets are reclustered using 
the Cambridge--Aachen algorithm~\cite{Dokshitzer:1997in,Wobisch:1998wt}.
The modified mass drop tagger algorithm~\cite{Dasgupta:2013ihk,Butterworth:2008iy}, also known as the soft-drop (SD) algorithm,
with an angular exponent $\beta = 0$, soft cutoff threshold $z_{\text{cut}} < 0.1$, and characteristic radius $R_{0} = 0.8$~\cite{Larkoski:2014wba}, is applied to remove soft and wide-angle radiation from the jet.

Jets containing \PQb hadrons (\PQb jets) are identified using a version of the combined secondary vertex algorithm based
on a deep neural network~\cite{csvtag,CMS-BTV-16-002}.
A working point with an identification efficiency of 70\% for genuine \PQb jets and a rejection of 99\% for light-flavor jets is used for AK4 jets. 
The \PQb tagging efficiencies in simulated event samples are corrected to match those measured in data.

The AK8 jets may be tagged as originating from a \PW boson decaying to $\PQq'\PAQq$ (\PW jet) or
from a top quark decaying into a fully hadronic final state (\PQt jet).
An AK8 jet is identified as a \PW jet by using the groomed SD mass $m_{\text{jet}}$, which
is required to be between 65 and 105\GeV. A requirement on the decorrelated $N$-subjettiness
ratio denoted as $\tau^{\mathrm{DDT}}_{21}$, which is constructed to decorrelate the
dependence on the jet \pt scale and mass~\cite{ddt, Sirunyan:2019jbg}, is also applied.
We first define $\tau_{21}=\tau_{2}/\tau_{1}$~\cite{Thaler:2010tr},
and then the decorrelated ratio is
$\tau^{\mathrm{DDT}}_{21}=\tau_{21} - M \ln[m^2_{\text{jet}}/(\pt^\text{jet} \mu)]$, 
where $\pt^\text{jet}$ is the jet \pt, $\mu = 1\GeV$, and $M$ is a parameter 
equal to $-0.080$ ($-0.082$) for 2016 (2017 and 2018) data.
The requirement $\tau^{\mathrm{DDT}}_{21} < 0.43$ is applied,
corresponding to an efficiency varying between 9 and 31\% over the kinematic range of this analysis for the combined $m_{\text{jet}}$ and $\tau^{\mathrm{DDT}}_{21}$ requirements, and an efficiency of incorrectly tagging a jet from a quark or gluon between 0.4 and 3.6\%, increasing with the jet \pt.
The variables for \PW tagging are calibrated in a \ttbar sample enriched in hadronically 
decaying $\PW$ bosons~\cite{Khachatryan:2014vla}. 

The combined secondary vertex algorithm for \PQb tagging is also applied to the subjets that constitute the internal structure
of an AK8 jet after the removal of the recognized soft radiation. In this case a looser working point, 
with an identification efficiency of 85\% for genuine \PQb jets and a rejection of 90\% for
light-flavor jets, is applied.

An AK8 jet is identified as a \PQt jet if it has a $\pt > 400\GeV$, at least one subjet \PQb tagged, 
a groomed SD mass between 105 and 220\GeV, and if it satisfies the requirement on the $N$-subjettiness ratio
$\tau_{32}=\tau_{3}/\tau_{2} < 0.65$.
A decorrelated $N$-subjettiness ratio has not been constructed in this case, but the overall performances of tagging criteria have been optimized targeting specifically the identification of top quarks.
The variables for top tagging are calibrated in a \ttbar-enriched sample~\cite{Sirunyan_2020}.
The combined SD mass and $\tau_{32}$ requirements correspond to 
a \PQt jet tagging efficiency of about 39\%, 
and an efficiency of incorrectly identifying a jet from another quark as a \PQt jet of between 1.8 and 3.5\%, 
depending on the data-taking year.

Electron and muon candidates are reconstructed in order to reject background events containing leptons.
Electron candidates are reconstructed in the fiducial region $\abs{\eta} < 2.5$ by matching the energy deposits in the ECAL 
with tracks reconstructed in the tracker~\cite{Khachatryan:2015hwa}. 
Electron identification criteria are based on the energy distribution in the ECAL, on the ratio 
of the energies measured in the HCAL and ECAL in the region around the electron candidate, 
and on the compatibility of the track with the primary vertex of event.
Muon candidates are reconstructed within the acceptance of the CMS muon system $\abs{\eta} < 2.4$ by using 
information from both the muon and tracker systems~\cite{Sirunyan:2018fpa}.
Muon identification criteria are based on the track impact parameter, the track reconstruction quality, and 
the number of hits observed in the tracker and muon systems.
Tracks of leptons (electrons and muons) are required to be isolated from unassociated energy deposits in the event. An upper threshold is applied 
to an isolation variable~\cite{Khachatryan:2015hwa, Sirunyan:2018fpa} defined as the scalar \pt sum of the charged and neutral hadrons and photons in a cone around the lepton direction of size $\Delta R = \sqrt{\smash[b]{ (\Delta \eta)^2 + (\Delta \phi)^2}}$,
where $\phi$ is the azimuthal angle in radians. The cone size is 0.4 for muons and 0.3 for electrons.
The sum is then corrected for pileup effects and divided by the lepton \pt. 

In the offline analysis, events are selected if they have at least one AK4 jet
with $\pt > 30\GeV$ and $\abs{\eta} < 4.0$, and $\ptmiss > 200\GeV$. 
In order to suppress multijet background events, the
angular separation $\Delta \phi_{\text{jet}, \ptvecmiss}$ between every AK4 jet
and the \ptvecmiss must be greater than 0.6.
Events containing identified electrons or muons are vetoed.

Three different top quark candidate reconstruction strategies are exploited.
A top quark candidate can be identified as a single \PQt jet (merged case), or it can be reconstructed from the
combination of one \PW jet and one \PQb jet (partially merged case), or of three AK4 jets (resolved case).

In the partially merged case, if multiple \PW jets are present in the event, the one with the highest \pt is chosen for the top quark reconstruction, and 
\PQb jets are considered for top quark reconstruction only if
they are separated from the \PW jet by $\Delta R_{\PW,\PQb} > 1.2$.
For the resolved top quark reconstruction, at least one of the three jets used for the combination 
must have been tagged as one \PQb jet.
In all three cases, if multiple top quark candidates of the same type are reconstructed, the one with the 
largest \pt is selected for the subsequent steps of the analysis.

Each selected event is assigned to one of six exclusive categories, defined according 
to the type of the reconstructed top quark candidate, \ie, merged, partially merged, or resolved case,
and to the presence or not of at least one forward jet.
If multiple types of top quark candidates are reconstructed, the event is assigned to only 
one category according to the following hierarchy, established after an optimization 
procedure aimed at achieving the best expected exclusion limit over the 
whole mass range: first the merged category, 
then the partially merged category, and finally the resolved category.
Additional requirements are applied to events in the resolved category 
in order to enhance the sensitivity: the resolved top quark candidate must have $\pt > 250\GeV$, 
and the event \HT, defined as the scalar sum of the AK4 jets \pt, must be greater than 200\GeV.

\begin{figure}[!htp]
\centering
\includegraphics[width=0.42\textwidth]{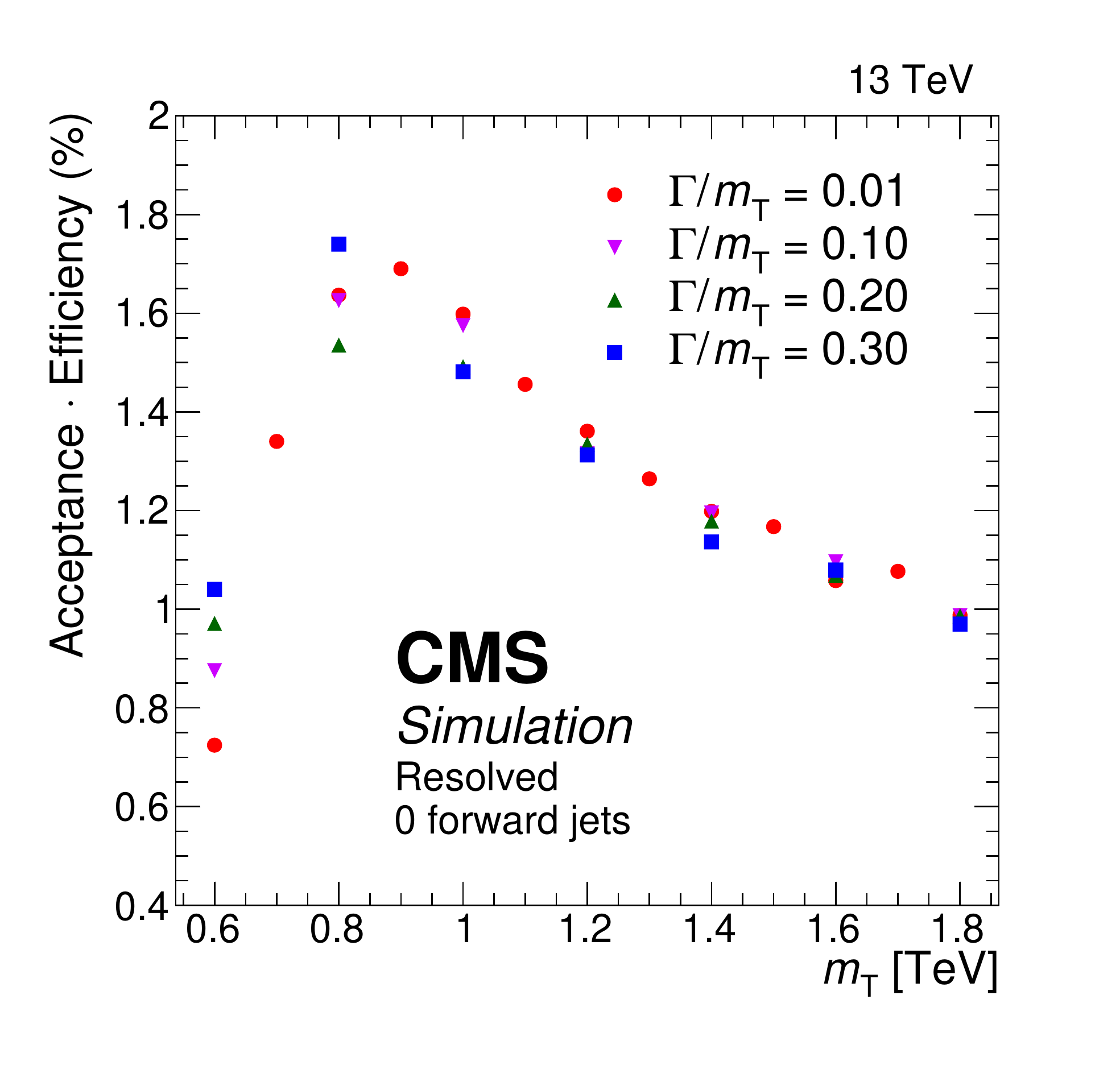}
\includegraphics[width=0.42\textwidth]{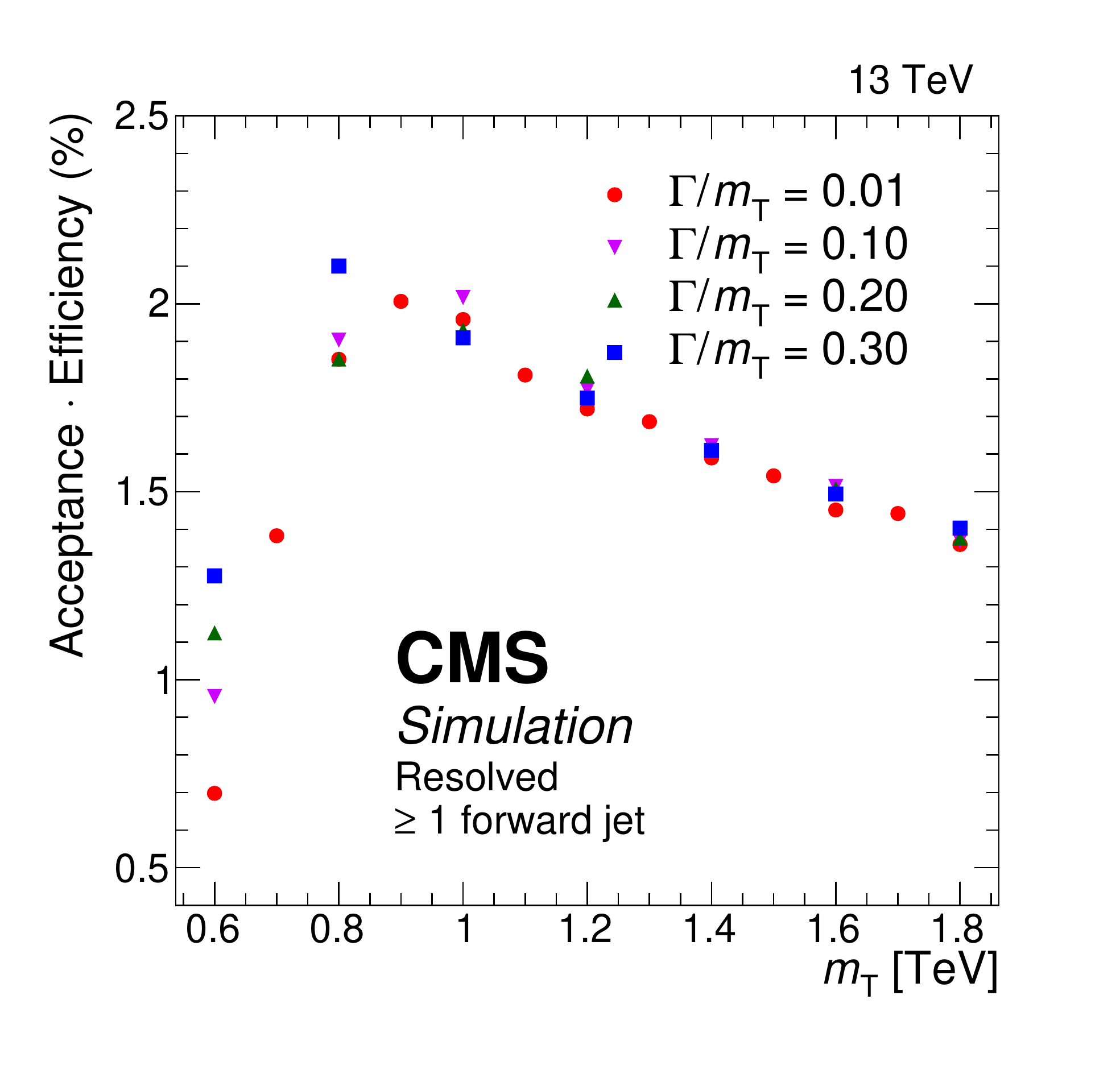}\\
\includegraphics[width=0.42\textwidth]{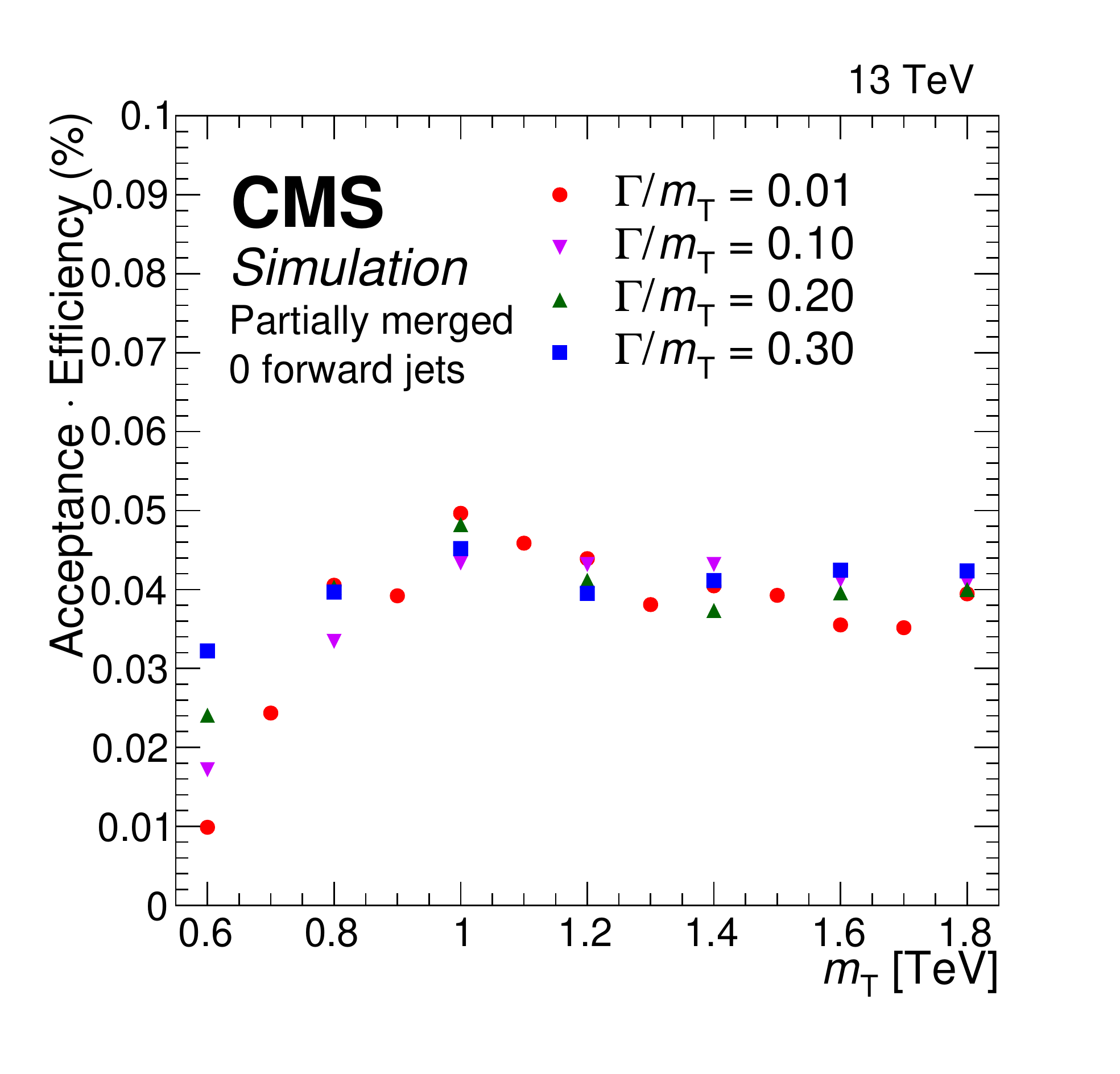}
\includegraphics[width=0.42\textwidth]{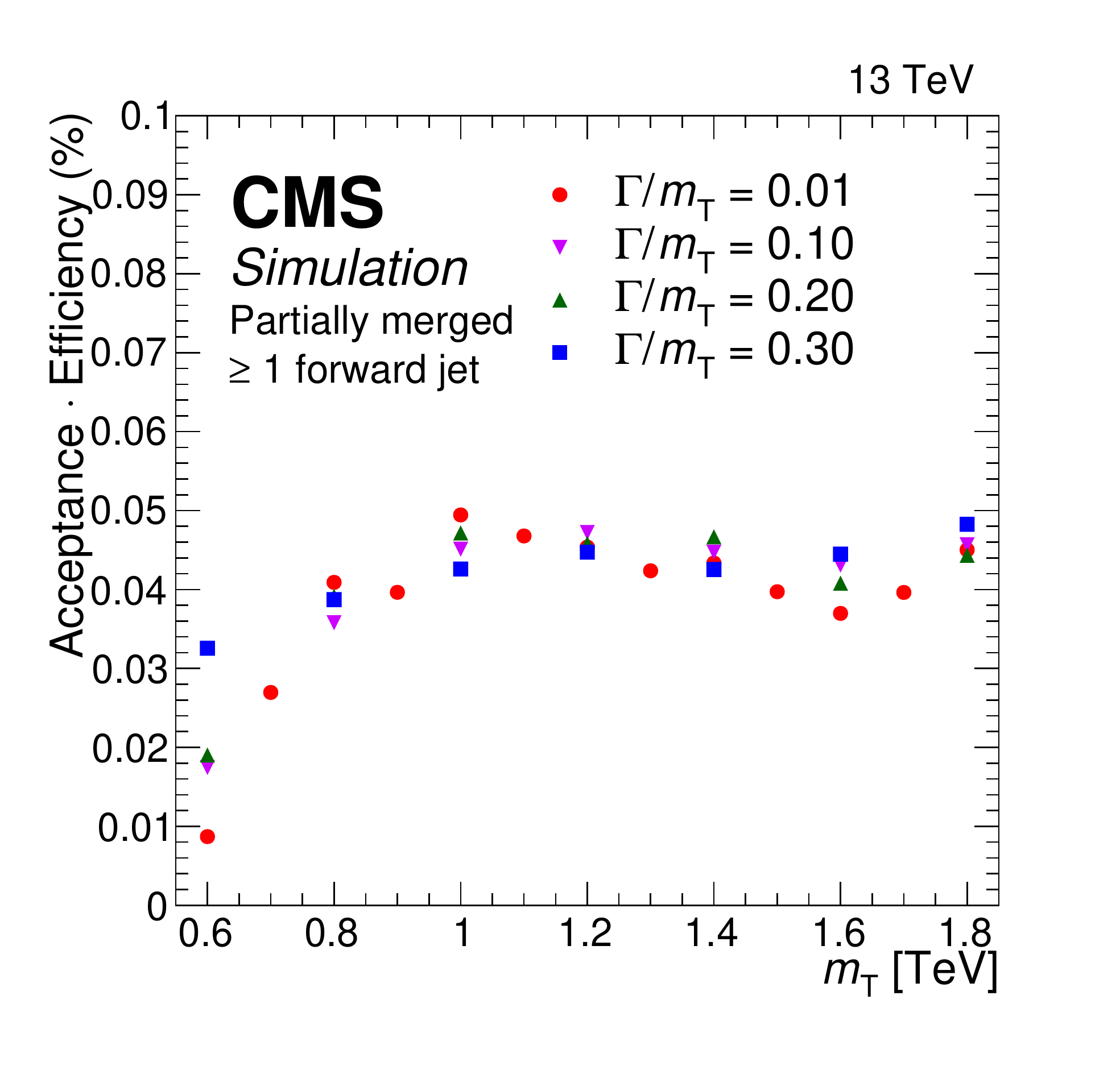}\\
\includegraphics[width=0.42\textwidth]{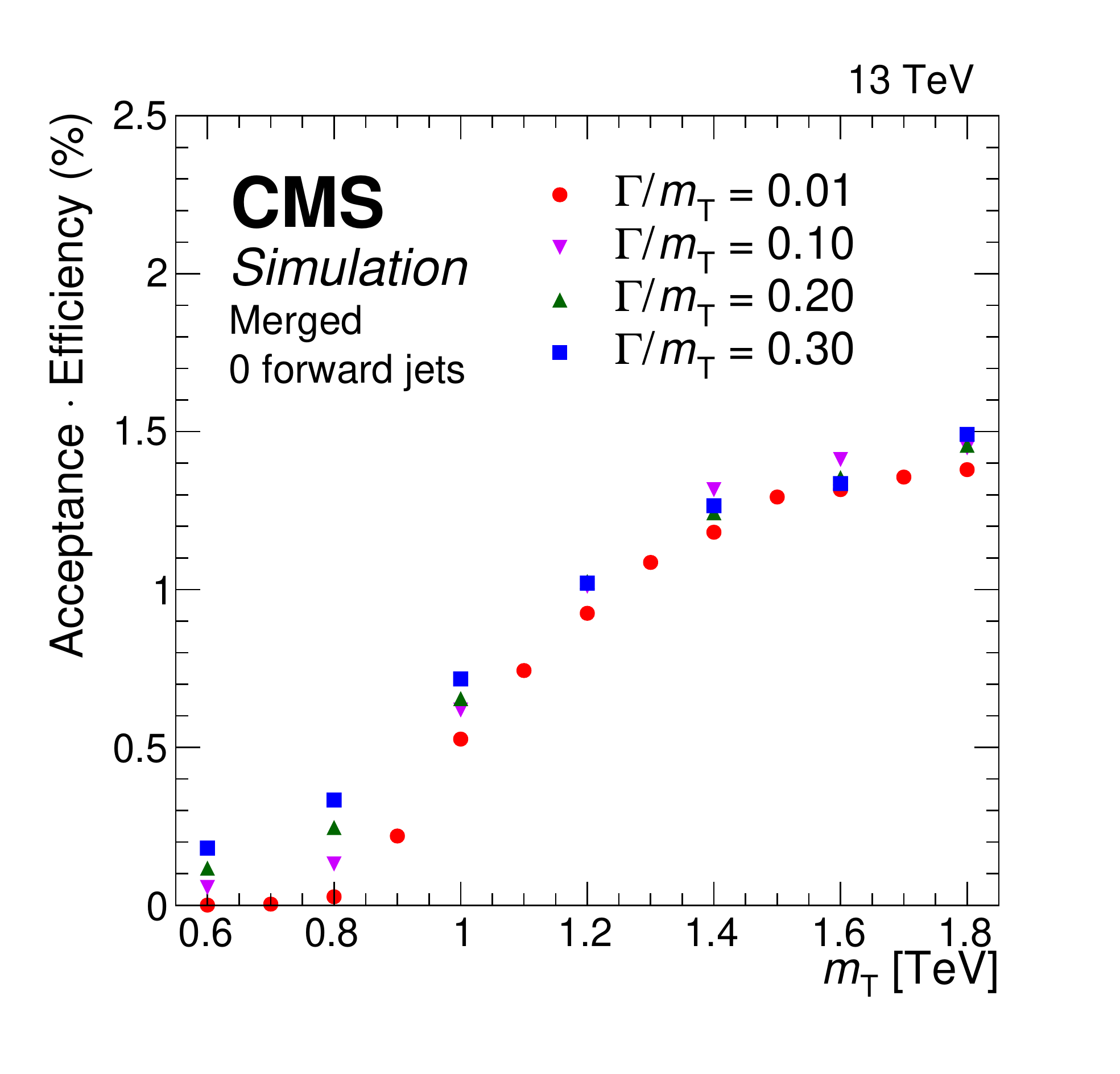}
\includegraphics[width=0.42\textwidth]{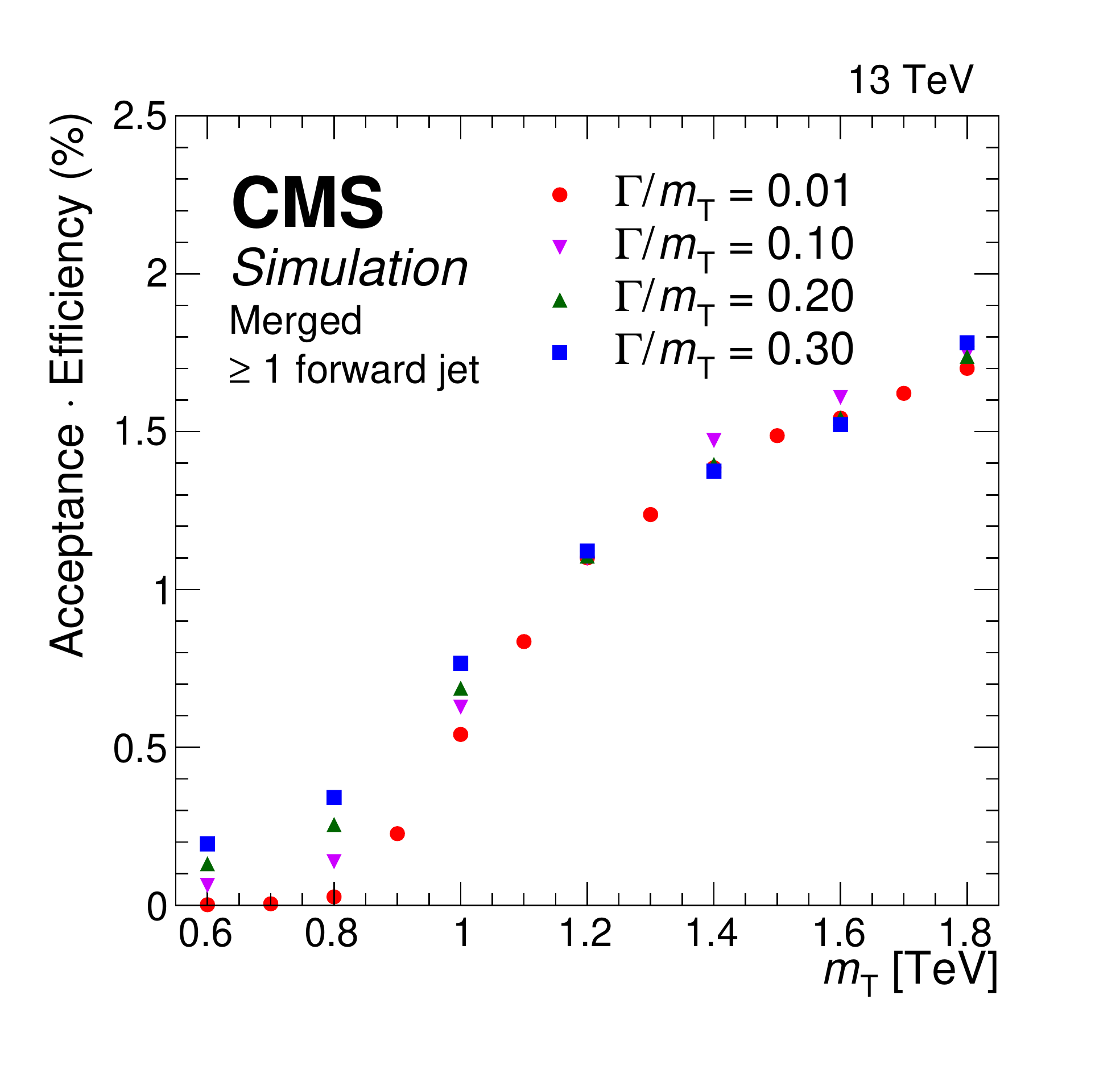}
\caption{The product of the signal acceptance and efficiency, averaged over the three data-taking years, for the final event selection in the resolved (upper), partially merged (central), and merged (lower) categories, for the zero forward jets (left), and at least one forward jet (right) categories. The numerator is the number of events passing the respective selection, the denominator is the total expected number of events. All possible decays of the top quark and \PZ boson are considered in both numerator and denominator of the efficiency computation, therefore the values are not corrected for the branching fraction of the final state.}
\label{fig:sigefficiencies}
\end{figure}

The signal selection efficiency, including both the contribution from detector acceptance and selection efficiency, in each of the six event categories is reported in Fig.~\ref{fig:sigefficiencies}, for all the masses and widths
of the \PQT quark considered in the analysis. All possible decays of the top quark and \PZ boson are considered. 
The efficiency is averaged over the three data-taking years according to the respective integrated luminosities.
The figure shows that, for the resolved analysis, the signal efficiency reaches the highest value for 
the mass hypotheses around 0.8\TeV after correction for threshold effects due to kinematic requirements applied, and decreases as the hypothesized resonance mass increases.
The opposite trend can be seen for the merged analysis, which in contrast displays an efficiency 
that increases with the mass hypothesis. 
This behavior is consistent with the top quark decay products being more collimated for higher top quark momentum regimes,
corresponding to higher \PQT quark mass hypotheses.
The partially merged analysis has a lower efficiency than the other two; nonetheless, it helps to recover signal events that are not selected by the merged and resolved analyses.  
The wide width interpretations feature similar efficiencies to the narrow width case, 
except for the lower mass points, where the broader mass distribution allows
more events to pass the momentum thresholds required in the selections.

The signal extraction is based on a simultaneous fit to the transverse mass
of the top quark candidate and \ptvecmiss system,
$\MT = \sqrt{\smash[b]{2 \pt^{\PQt} \ptmiss (1-\cos\Delta\phi_{\PQt, \ptvecmiss})}}$,
in the six analysis categories.

\section{Background estimation}\label{sec:background}
The major sources of background after the event selection are \ttjets, \wjets, and \zjets events where the \PZ boson decays to neutrinos.
For the merged categories the \zjets background is the largest, followed by significant contributions from \wjets and \ttjets events.
For the partially merged categories the \ttjets background is dominant.
For the resolved categories the \ttjets background is the leading one, followed by significant contributions from \wjets and \zjets events. 
The contributions from minor backgrounds, such as single top quark and multijet events, are estimated from simulation, since their contribution is smaller than the other two by more than one order of magnitude.

Corrections to the simulations are derived from dedicated background-enriched samples in data, referred to as control regions, and then applied to the background transverse mass (\MT) distributions in the signal regions before performing the fit.

Different strategies for the determination of the correction factors are employed for the resolved, the merged, and partially merged categories,
because the amount of data in the control regions and the mismodeling corrections are different.

For events in the resolved category, possible discrepancies between data and simulation distributions 
of the discriminating variables are likely to be due to unaccounted higher-order 
corrections in perturbative QCD or quantum electrodynamics calculations.
In this case,  the large numbers of events in the control regions allow the correction factors to be determined as functions of \HT and \ptmiss.

For events in the merged and partially merged categories, background processes can satisfy the selection requirements of the signal region because of jets that occasionally satisfy the requirements to be identified as \PQt and \PW jets. For such events data and simulation can differ because of fluctuations in the hadronic jets background and the lack of higher-order corrections in the simulation. In the signal regions, such jets are mostly produced in association with \PW or \PZ bosons. In this case, the correction factors are determined in ranges of the observable \MT, and a smoothing procedure is applied.

Independent sets of control regions are defined for the resolved, the merged, and the partially merged categories.
Starting from the baseline selection of each category, one or more requirements are inverted in order to obtain
a properly background-enriched sample of events.
In the following, only the requirements that have been inverted with respect to the signal selection 
for each category are described, while all the others stay the same.

In the resolved category, the \ttbar control region is defined by requiring at least one identified lepton in the event. 
The {\PW}+jets control region is defined by selecting events with at least one identified lepton and no \PQb jets.
The {\PZ}+jets control region is defined by selecting events with no \PQb jets.

In the merged category, the \ttbar control region is defined by requiring at least one identified lepton in the event
and removing the requirement on the minimum $\Delta \phi_{\text{jet}, \ptvecmiss}$ from the selection. 
A single control region is defined for \zjets and \wjets by requiring that no subjet of the merged top quark candidate has been identified as one \PQb jet.

In the partially merged category, the background is largely given 
by \ttbar events, therefore only a control region for \ttbar is used,
which is defined by requiring the minimum $\Delta \phi_{\text{jet},  \ptvecmiss} < 0.6$. 

\begin{figure}[htbp!]
\centering
\includegraphics[width=0.38\textwidth]{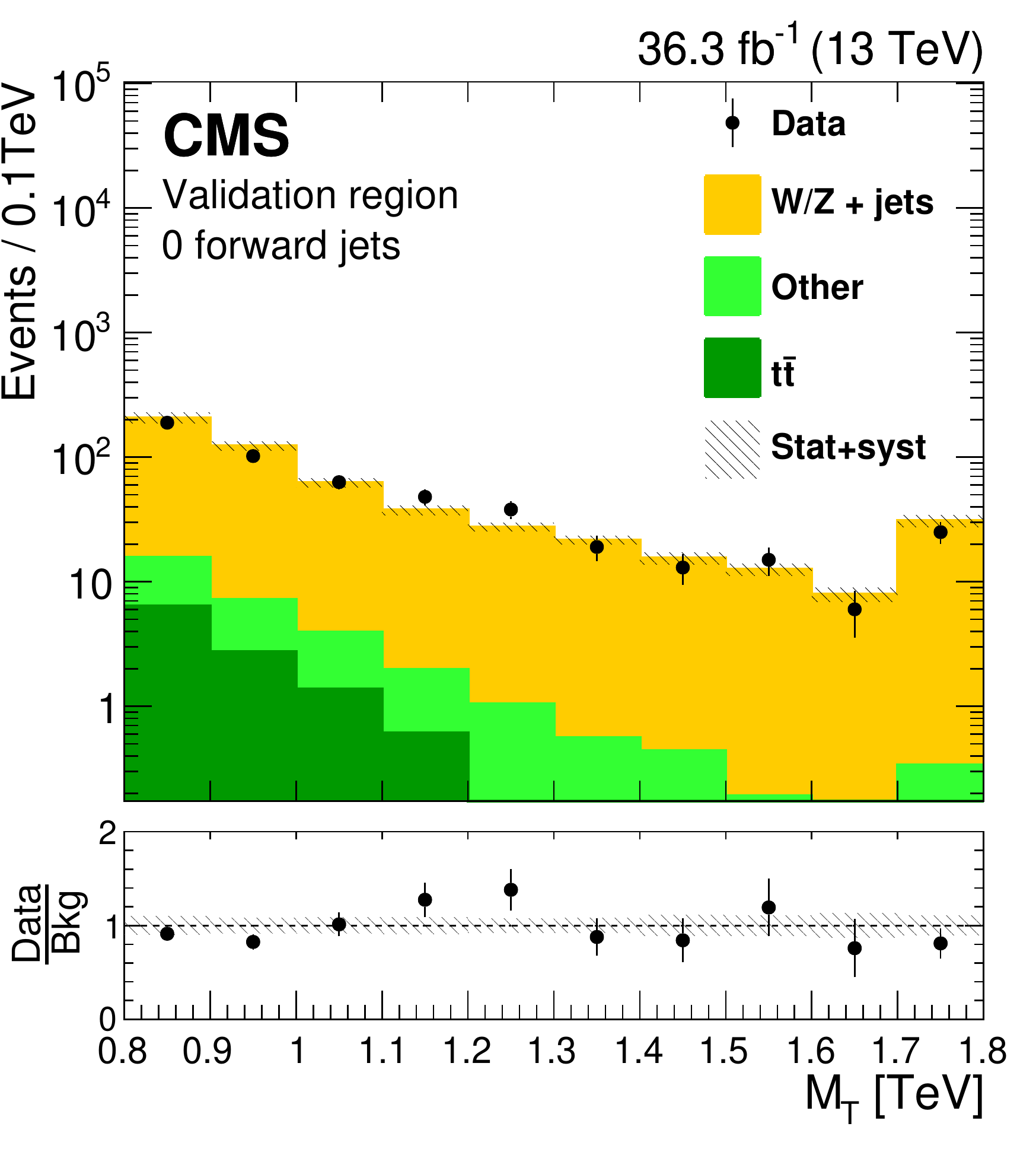}
\includegraphics[width=0.38\textwidth]{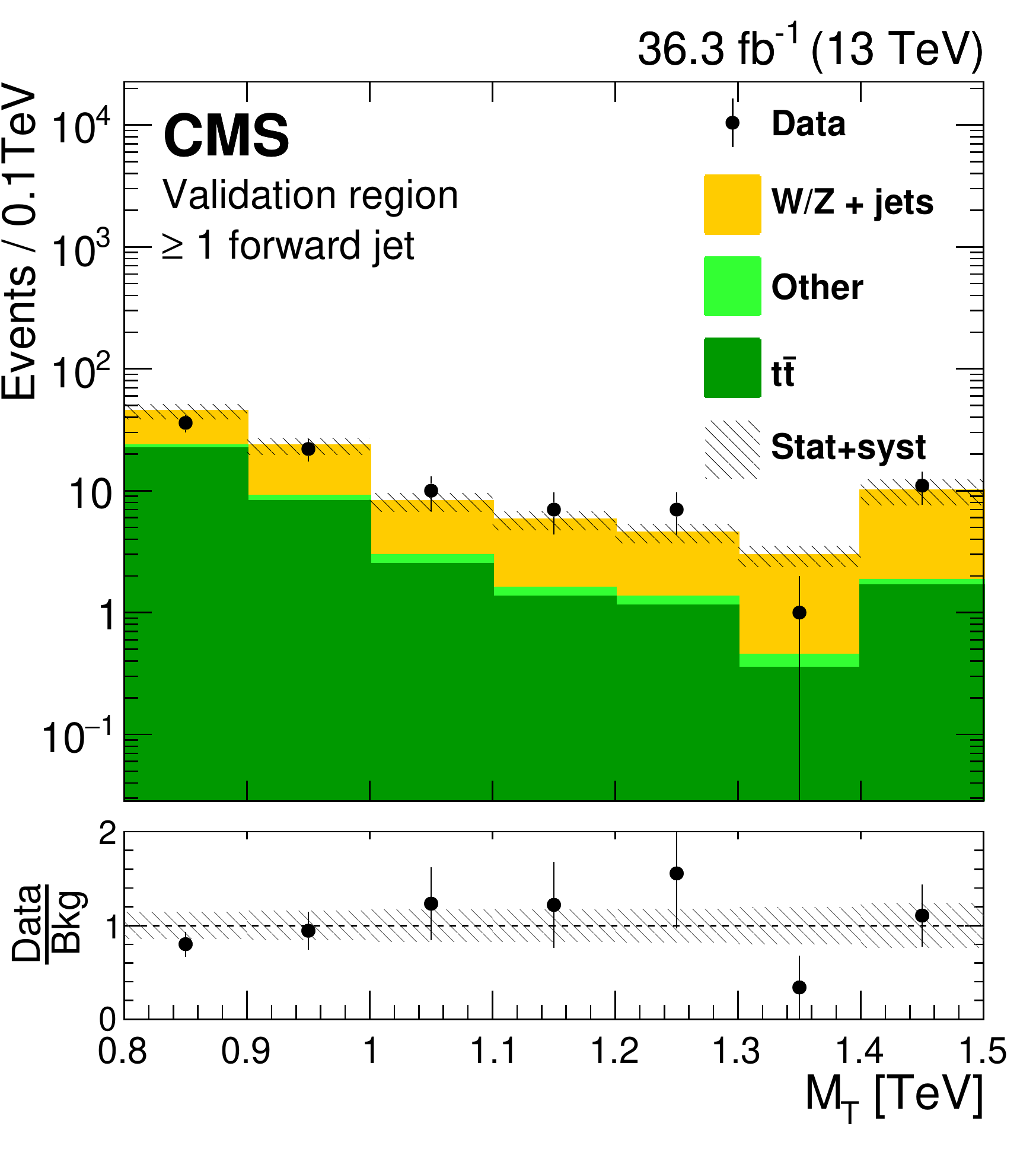}\\
\includegraphics[width=0.38\textwidth]{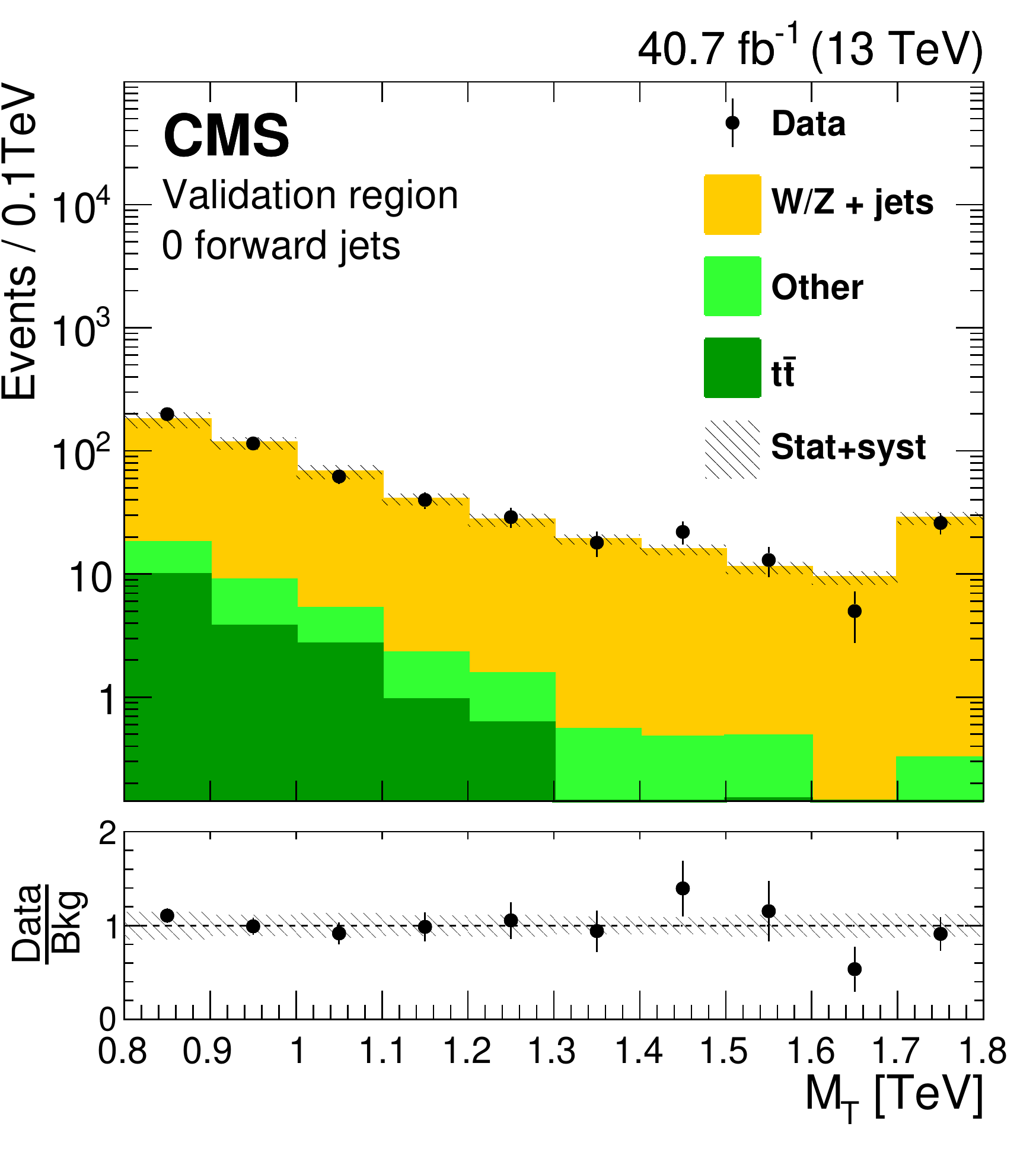}
\includegraphics[width=0.38\textwidth]{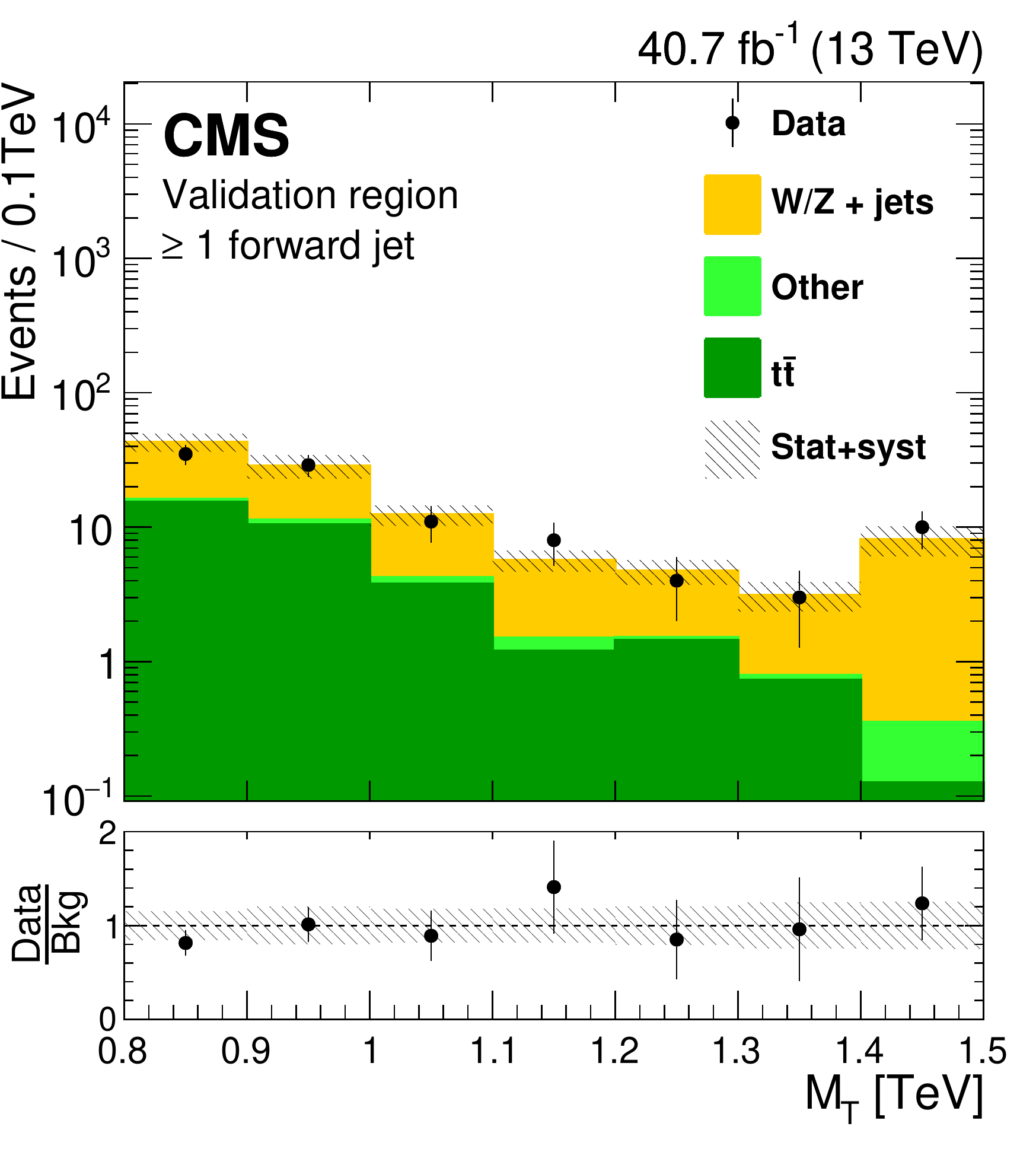}\\
\includegraphics[width=0.38\textwidth]{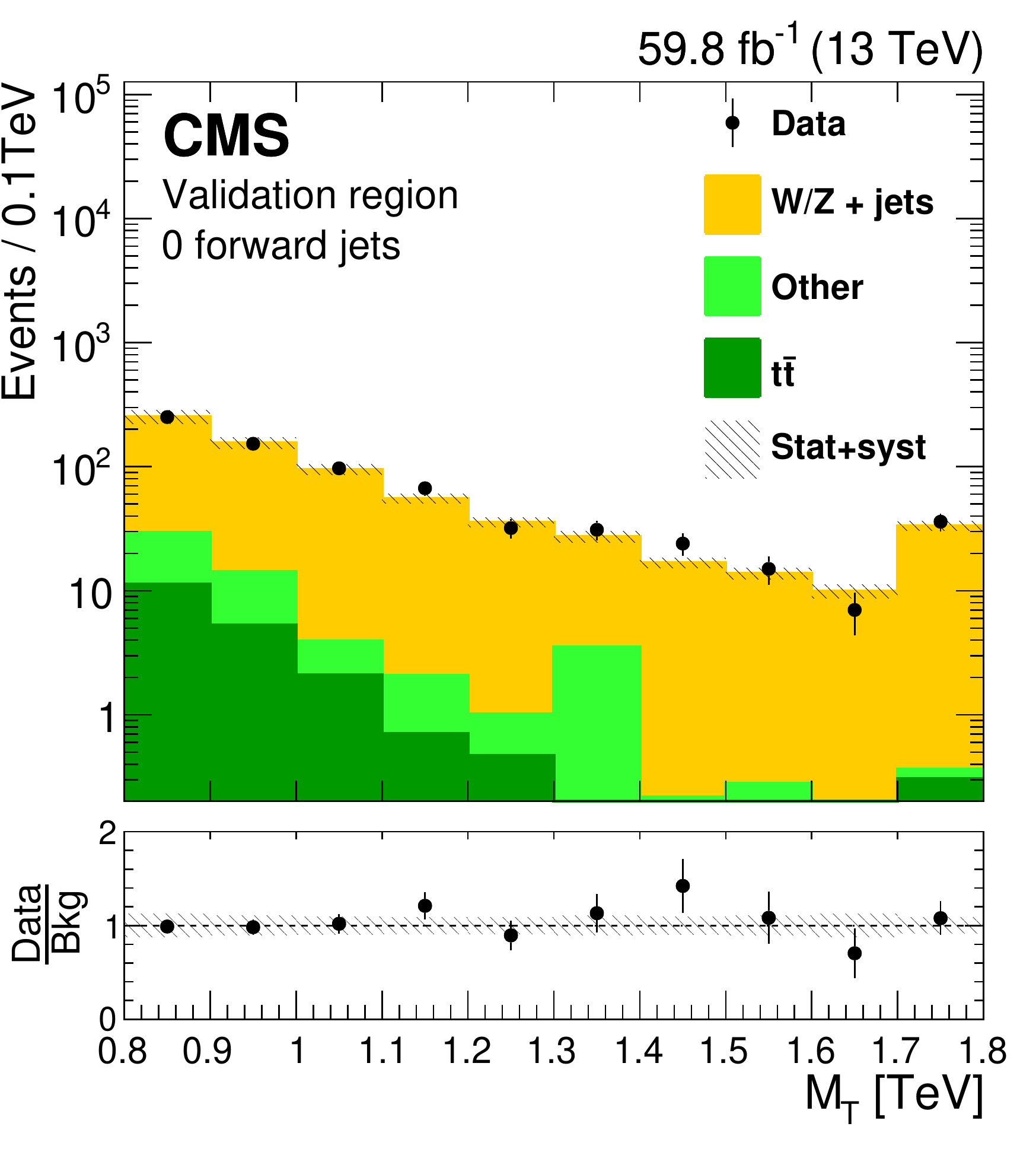}
\includegraphics[width=0.38\textwidth]{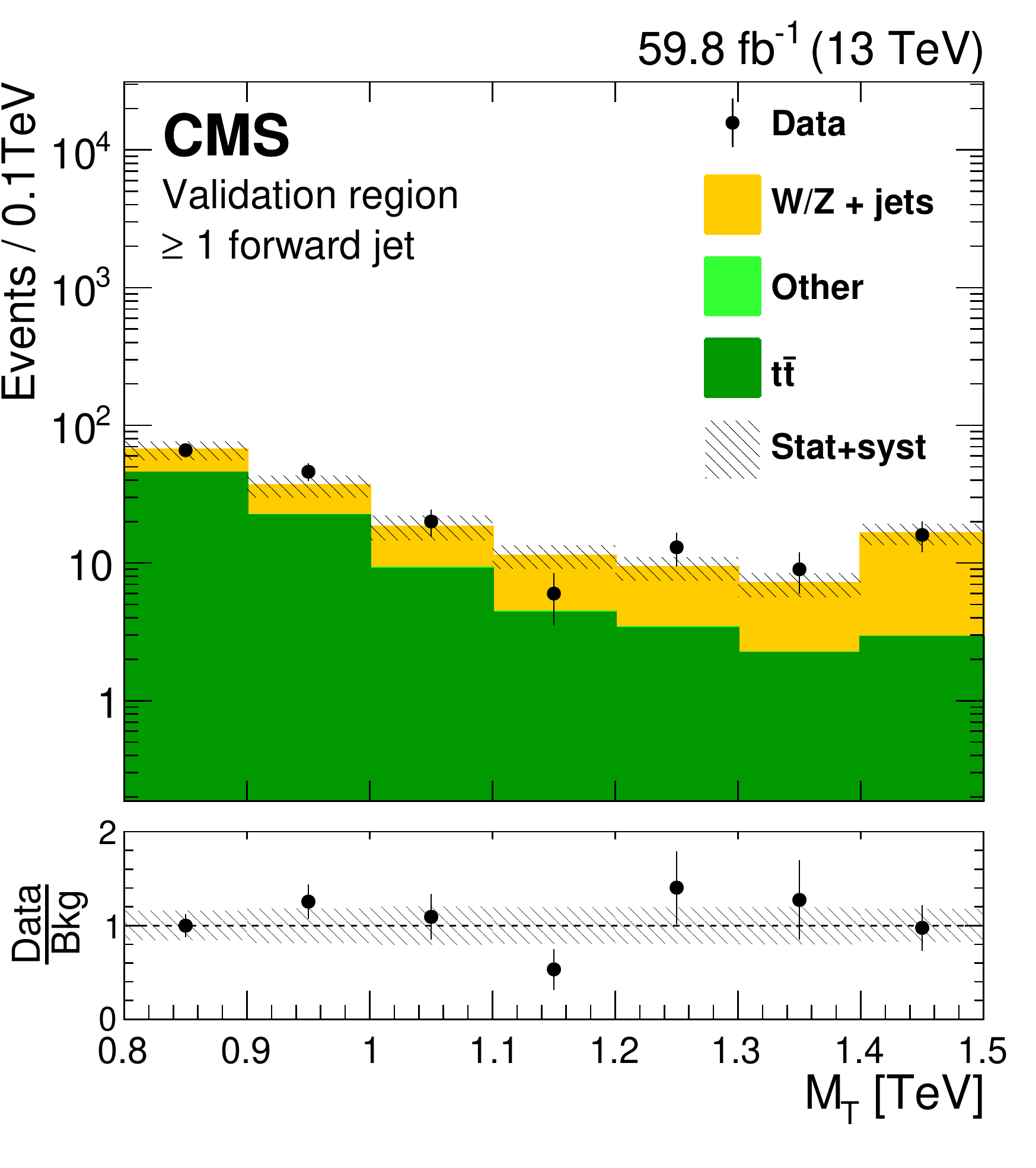}
\caption{Distributions of the transverse mass \MT of the reconstructed top quark and \ptvecmiss system, for events in validation samples selected as in the merged categories, but with the AK8 jet SD mass outside the interval 105--220\GeV, for events with no forward jet (left) and at least one forward jet (right), and for 2016 (upper), 2017 (central), and 2018 (lower). The overflow is included in the last bin. The predictions of the main background components have been determined in simulation with scale factors applied to match data extracted from control regions. The uncertainties after the fit to data in the validation regions are shown.}
\label{fig:MtMergedCheck}
\end{figure}

\begin{figure}[htbp!]
\centering
\includegraphics[width=0.4\textwidth]{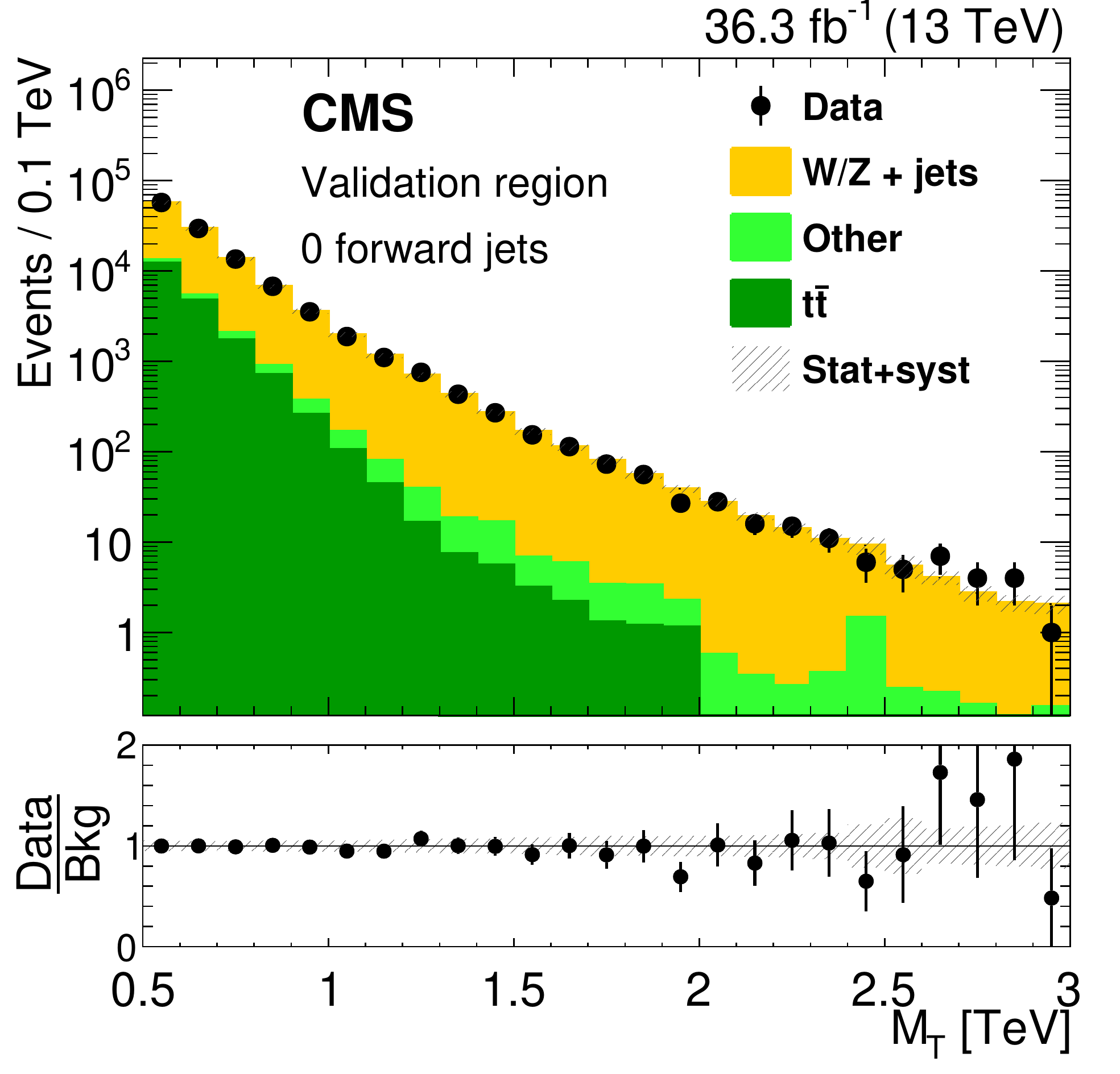}
\includegraphics[width=0.4\textwidth]{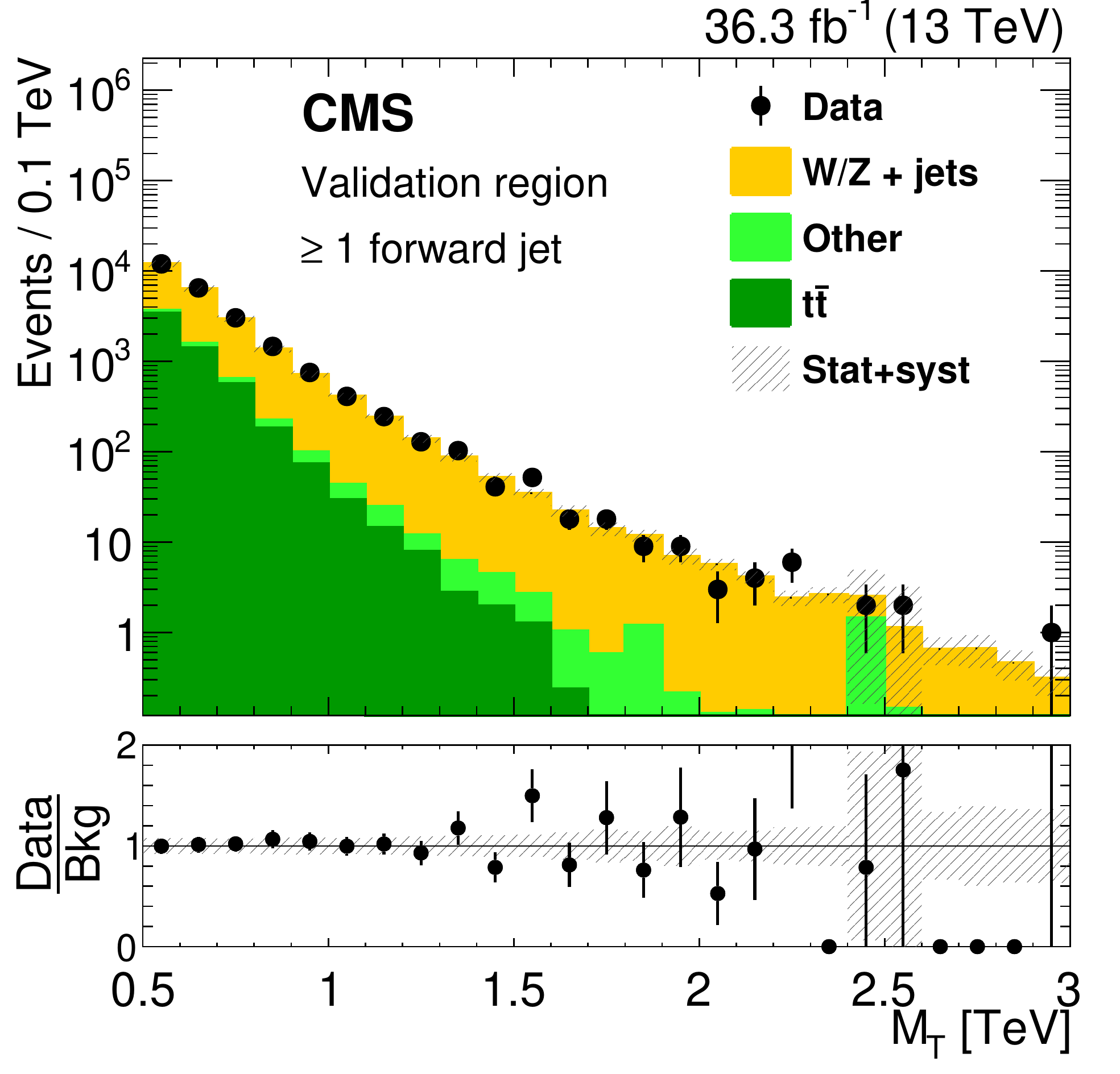}\\
\includegraphics[width=0.4\textwidth]{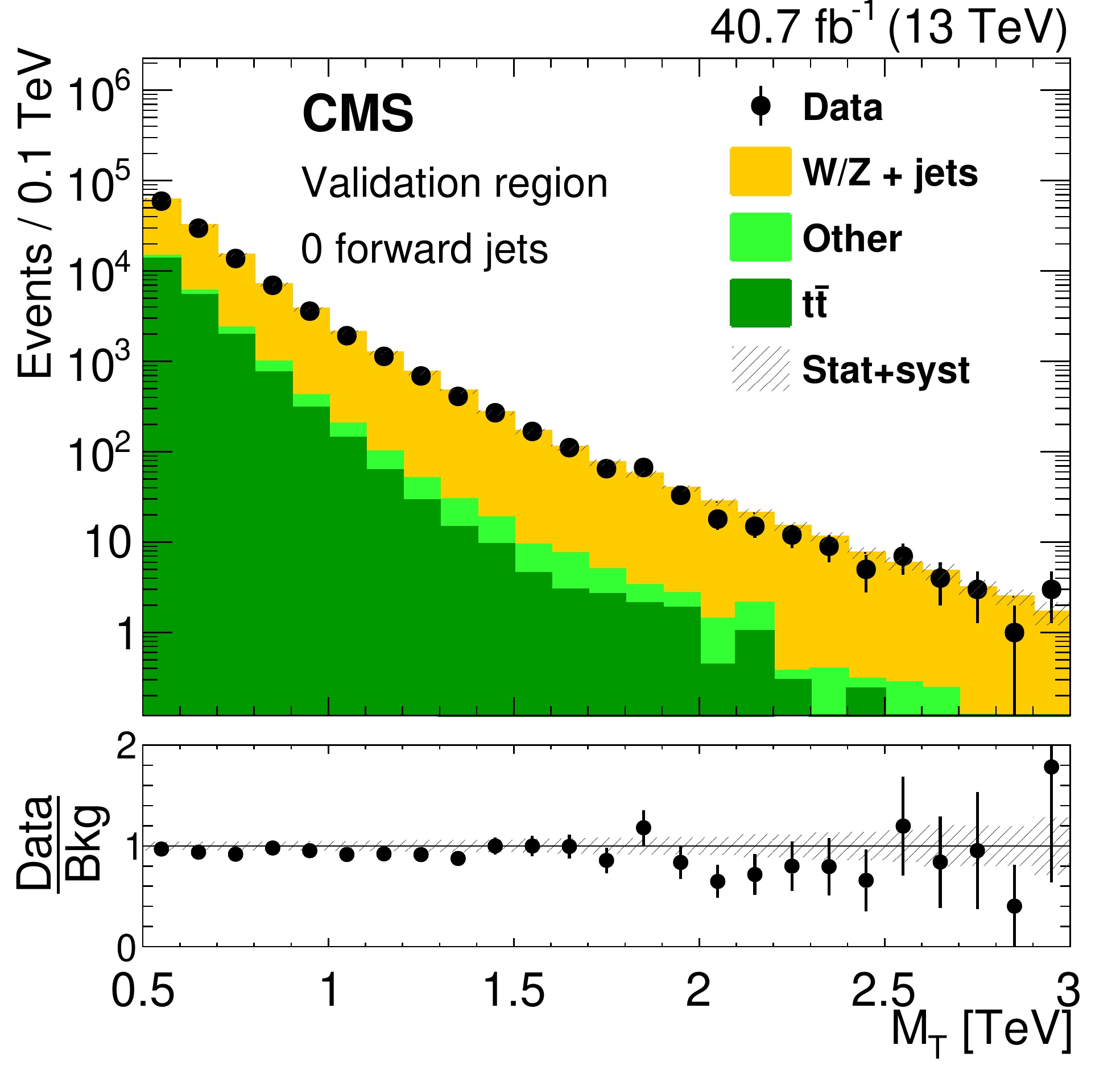}
\includegraphics[width=0.4\textwidth]{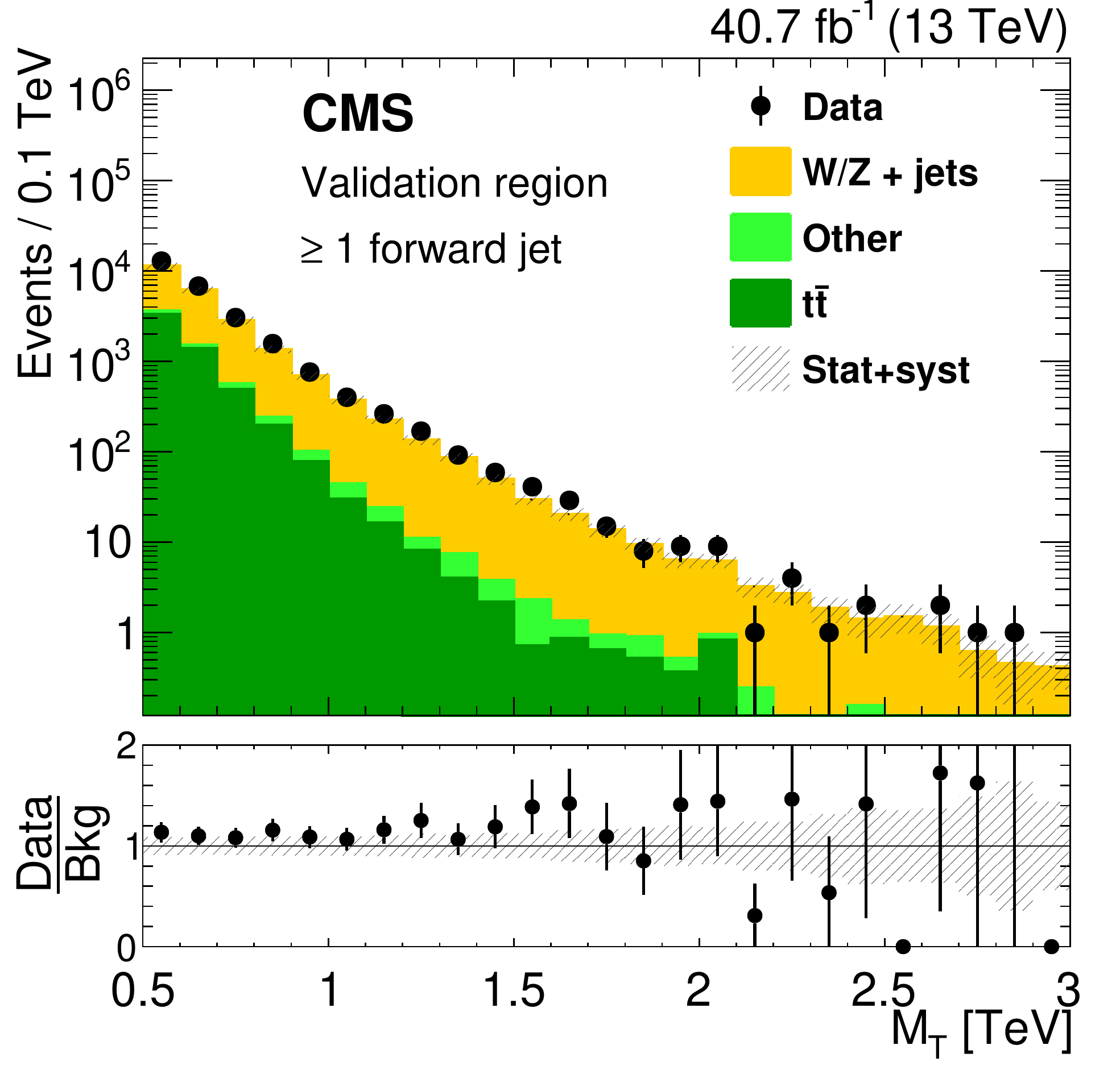}\\
\includegraphics[width=0.4\textwidth]{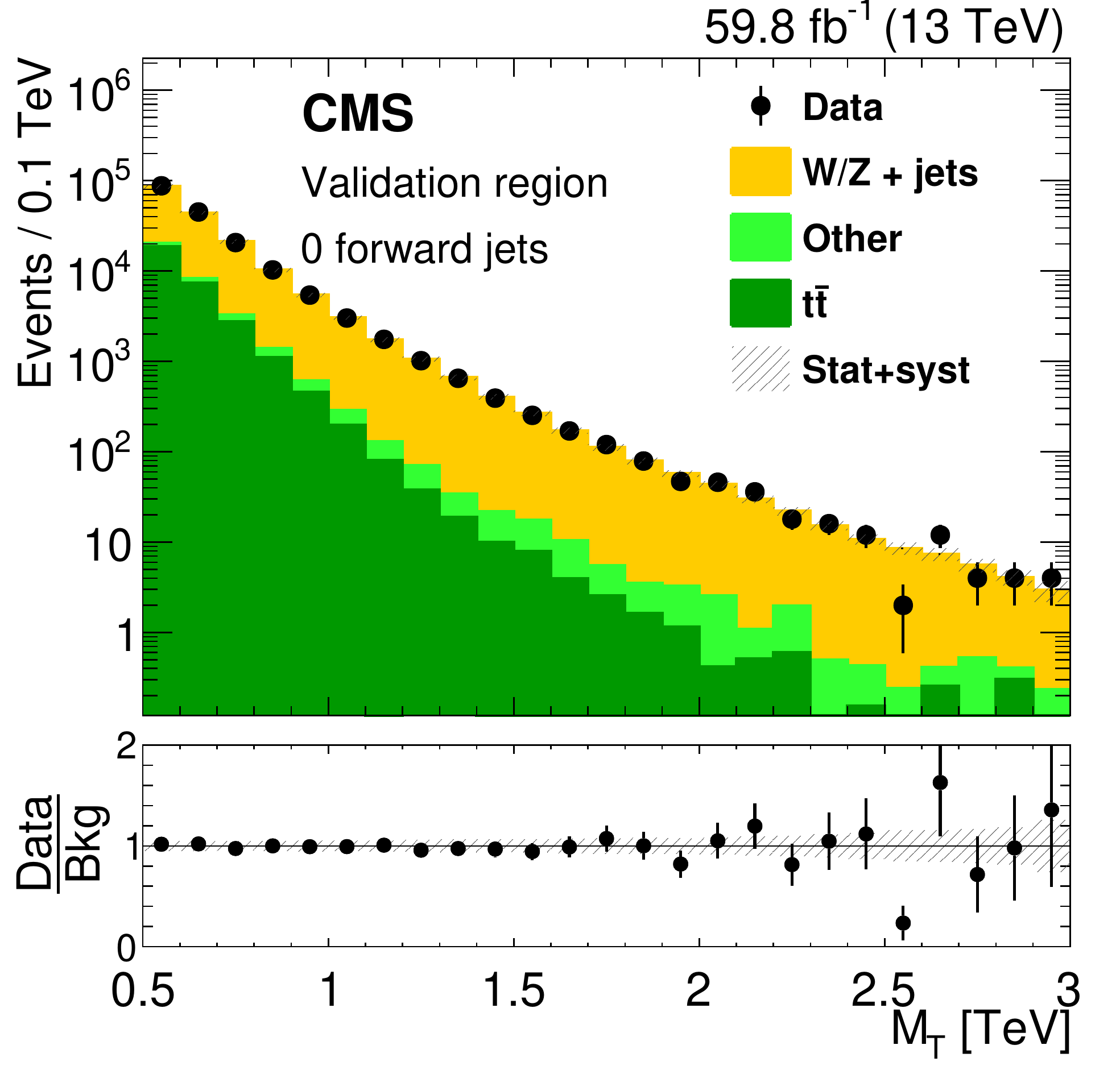}
\includegraphics[width=0.4\textwidth]{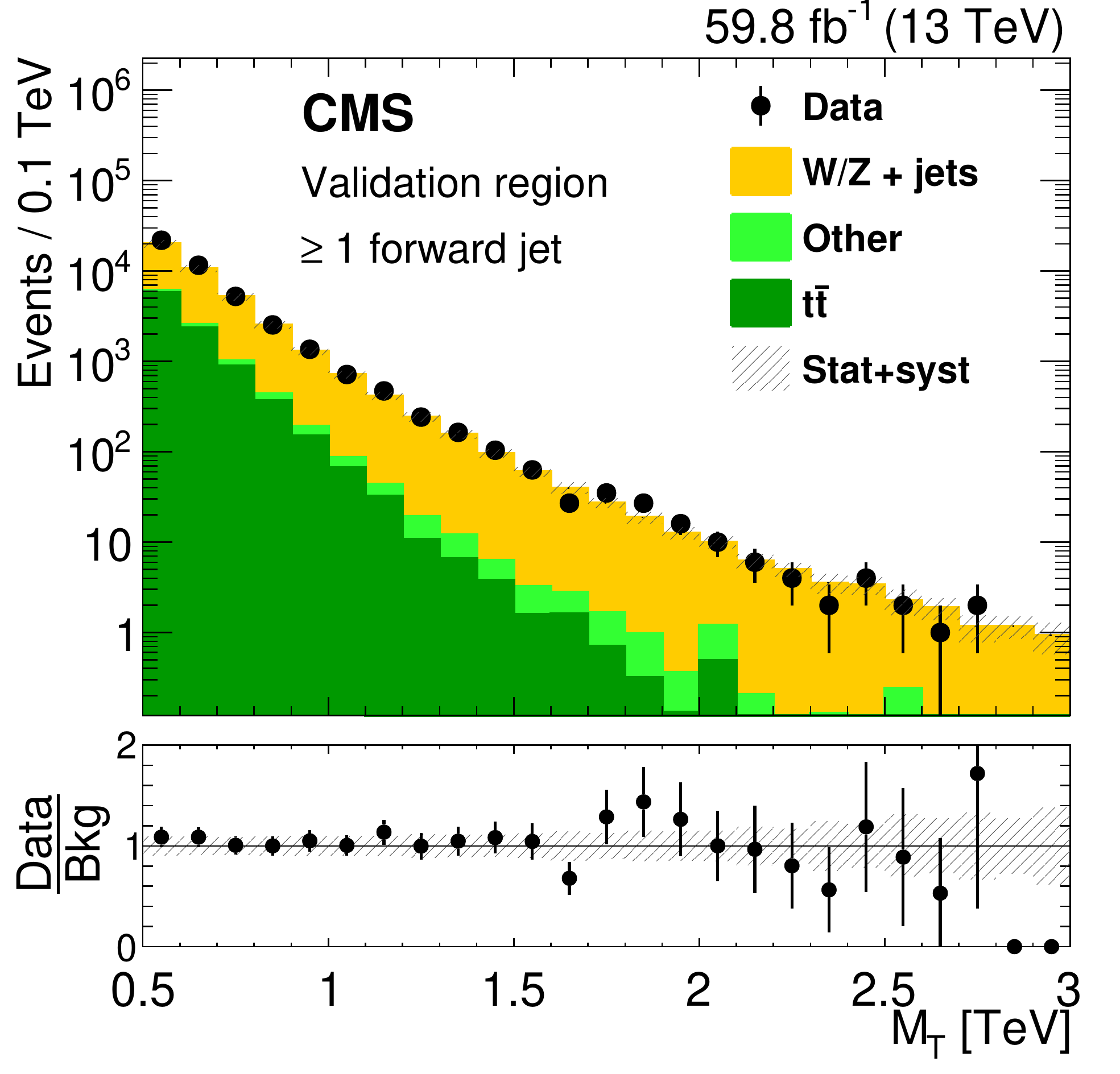}
\caption{Distributions of the transverse mass \MT of the reconstructed top quark and \ptvecmiss system, for events in validation samples selected as in the resolved categories, but without the requirement of at least one \PQb jet in the event, for events with no forward jet (left) and at least one forward jet (right), and for 2016 (upper), 2017 (central), and 2018 (lower). The predictions of the main background components have been determined in simulation with scale factors applied to match data extracted from control regions. The uncertainties before the fit to data in the validation regions are shown.}
\label{fig:MtResolvedCheck}
\end{figure}

The strategy employed for the background prediction in the merged and partially merged categories
is tested in an independent sample, defined with the same selection as the merged category but 
with the AK8 jet SD mass being outside the interval 105--220\GeV. 
For the resolved categories, the background estimation method is validated in a sample obtained from
a looser selection, defined by removing the requirement on at least one \PQb jet in the event. 
Figures~\ref{fig:MtMergedCheck} and~\ref{fig:MtResolvedCheck} show the comparison of data and the predicted backgrounds
in the above validation samples for the merged and resolved categories, respectively.

\section{Systematic uncertainties}\label{sec:systematics}
Systematic uncertainties, both of theoretical and of experimental origin, affect the background and signal prediction in terms of the pre-fit event rate, or the shape of the transverse mass (\MT) distributions, or both. Most of the uncertainties are related to corrections that are applied to simulated events in order to account for the known discrepancies with data. The background determination procedure described in Section~\ref{sec:background}, including the derivation of the correction factors, is repeated for each systematic uncertainty variation in order to derive the corresponding modified background \MT distributions.
A summary of the systematic uncertainties and their effect in the analysis is reported in Table~\ref{tab:systematics}.

The total 2016--2018 integrated luminosity has an uncertainty of 1.6\%~\cite{CMS-LUM-17-003}, which applies to the overall event rates of all simulated processes. 
 
The uncertainty associated with the mismodeling of pileup is evaluated by varying the $\Pp\Pp$ total inelastic cross section used to compute the pileup distribution in data by $\pm$4.6\%~\cite{Sirunyan:2018nqx}, resulting in a change in the normalization in the range 0.2--3\%, depending on the category, the year, and the signal or background component. 

Systematic uncertainties due to differences in the trigger efficiency between data and simulation are found to be 1--3\%.
During the 2016 and 2017 data-taking, a gradual shift in the timing of the inputs of the ECAL L1 trigger in the region $\abs{\eta} > 2.0$ caused a specific trigger inefficiency. For events containing a jet with $\pt \gtrsim 100\GeV$ in the region $2.5 < \abs{\eta} < 3.0$, the efficiency loss is approximately 10--20\%, depending on \pt, $\eta$, and data-taking period~\cite{Sirunyan:2020zal}. Correction factors were computed from data and applied to the acceptance evaluated by simulation, resulting in a systematic uncertainty of 0.2--3\%. 

The jet four-momenta are varied according to the jet energy scale and resolution uncertainties, resulting in a systematic uncertainty of 2--18 and 1--5\%, respectively. 
The uncertainty in the \PQb tagging efficiency for the \PQb, \PQc, and light-flavor quarks or gluon has been evaluated by varying the corrections applied to simulation by their uncertainties, giving a change in the normalization in the range 0.5--1.2\%. A similar procedure is done with the t and W jet tagging uncertainties, giving a variation in the normalization of 9--10 and 7--8\%, respectively. 

The uncertainties from the choice of the PDF are evaluated by reweighting the simulated signal and background events using the NNPDF~3.0 or 3.1 sets of eigenvectors~\cite{Butterworth:2015oua}. 

The choice of the factorization and renormalization scales, $\mu_\mathrm{F}$ and $\mu_\mathrm{R}$, affects both the cross section and the acceptance of each background. The uncertainties are estimated separately for each background by halving or doubling both scales together, compared to the nominal values used in simulation. The changes in the background rates are found to be between 10--30\%. For the signal this source of uncertainty affects only the acceptance.

The uncertainties associated with the PDF choice, the \PQb tagging efficiencies and mistag rates, and the pileup simulation are found to have negligible effects on the shape of the \MT distribution, so only their effects on the overall event rates are considered.

For each background contribution, the statistical uncertainty in each bin of the \MT distribution is taken into account.
All uncertainties are considered fully correlated across the three years of data-taking, except for those corresponding to the ECAL L1 trigger inefficiency and the background scale factors.

The effect on the background \MT distribution of the statistical uncertainties in the scale factors, 
due to the limited number of events in the data control regions and in simulation, is treated as follows: 
first the statistical uncertainty in the scale factors is estimated, 
then two possible uncertainties in the overall background distribution are considered. 
The first uncertainty considers the envelope of all bin variations, thus all scale factors 
variations are treated as fully correlated over each bin of the \MT spectrum. 
The second uncertainty considers the scale factor variation in each bin of \MT to linearly decrease
 from the maximum value at one extreme of the \MT spectrum to the minimum value at the other. 
This uncertainty corresponds to the maximum variation in the \MT slope allowed in data, and it is not uniform over the \MT spectrum, because the statistical uncertainty grows larger at the higher end of the spectrum. All these uncertainties are independently derived for each year and fit simultaneously to data by employing separate nuisance parameters.
This parametrization has been chosen to reduce the dependency on bin-by-bin statistical fluctuations, which may cause a bias on the signal yield extracted in the statistical procedure. The number of nuisance parameters considered is the minimum that allows closure in data validation regions, showing that the respective variation encompass the statistical variation seen in data.

\begin{table}[!ht]
\topcaption{Summary of the systematic uncertainties. The maximum range of change in the pre-fit event rate of signals and backgrounds across all years and categories for one standard deviation change in the systematic uncertainty is reported in the ``Effect'' column. All uncertainties are considered fully correlated across the three years of data-taking, except for those corresponding to the ECAL L1 trigger inefficiency and the background scale factors. The third column indicates whether the uncertainty affects both the rate and the shape of the distributions or the rate only. Except for the background scale factors, all the uncertainties affect both signal and background inputs to the fit.}\label{tab:systematics}
\centering
\begin{tabular}{l c c}
Source & Effect (\%) & Type\\
\hline
Integrated luminosity       & 1.8  & rate \\
Pileup           & 0.2--3    & rate \\
b tagging        & 0.5--1.2  & rate\\
t tagging      & 9--10    & rate, shape\\
W tagging        & 7--8      & rate, shape\\
Trigger efficiency & 1--3    & rate, shape\\
ECAL L1 trigger inefficiency & 0.2--3 & rate, shape\\
Jet energy scale & 2--18     & rate, shape\\
Jet energy resolution & 2--5 & rate, shape\\
PDF & 1--5 & rate\\
$\mu_\mathrm{R}$, $\mu_\mathrm{F}$ & 10--30 & rate, shape\\
Background scale factors & 5--50 & rate, shape\\[\cmsTabSkip]
\end{tabular}
\end{table} 

\section{Results}\label{sec:results}
\begin{figure}[htbp!]
\centering
\includegraphics[width=0.36\textwidth]{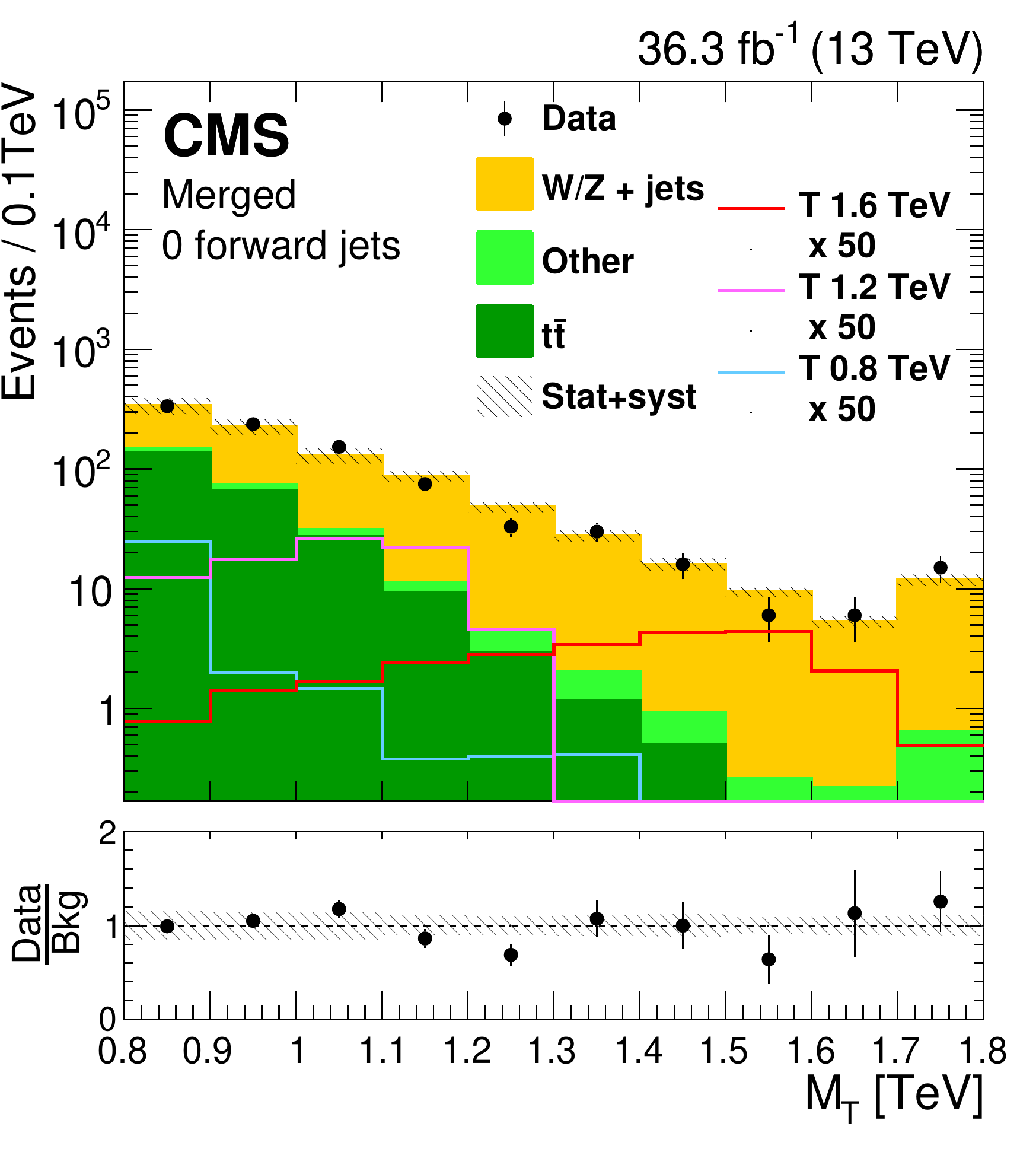}
\includegraphics[width=0.36\textwidth]{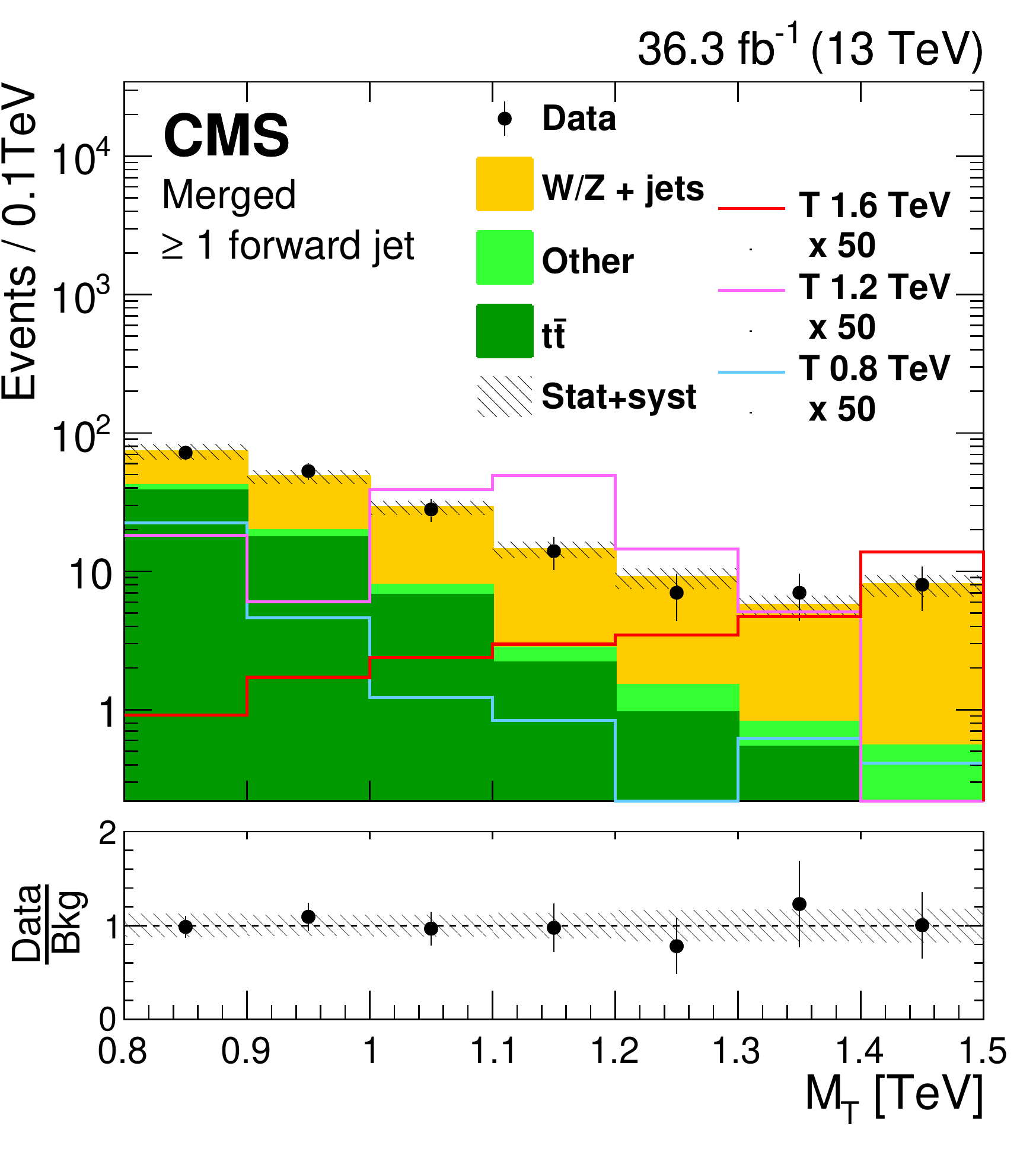}\\
\includegraphics[width=0.36\textwidth]{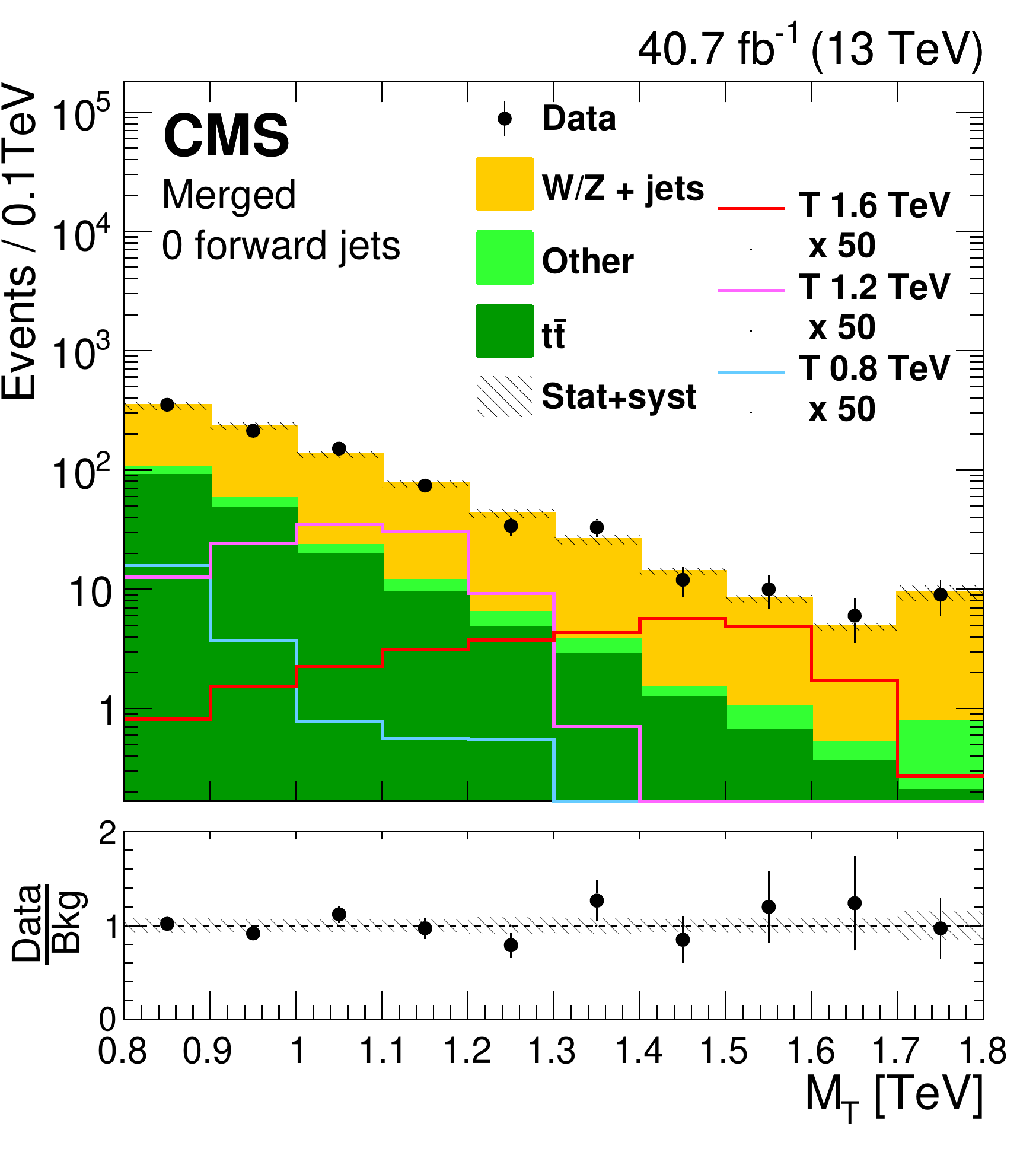}
\includegraphics[width=0.36\textwidth]{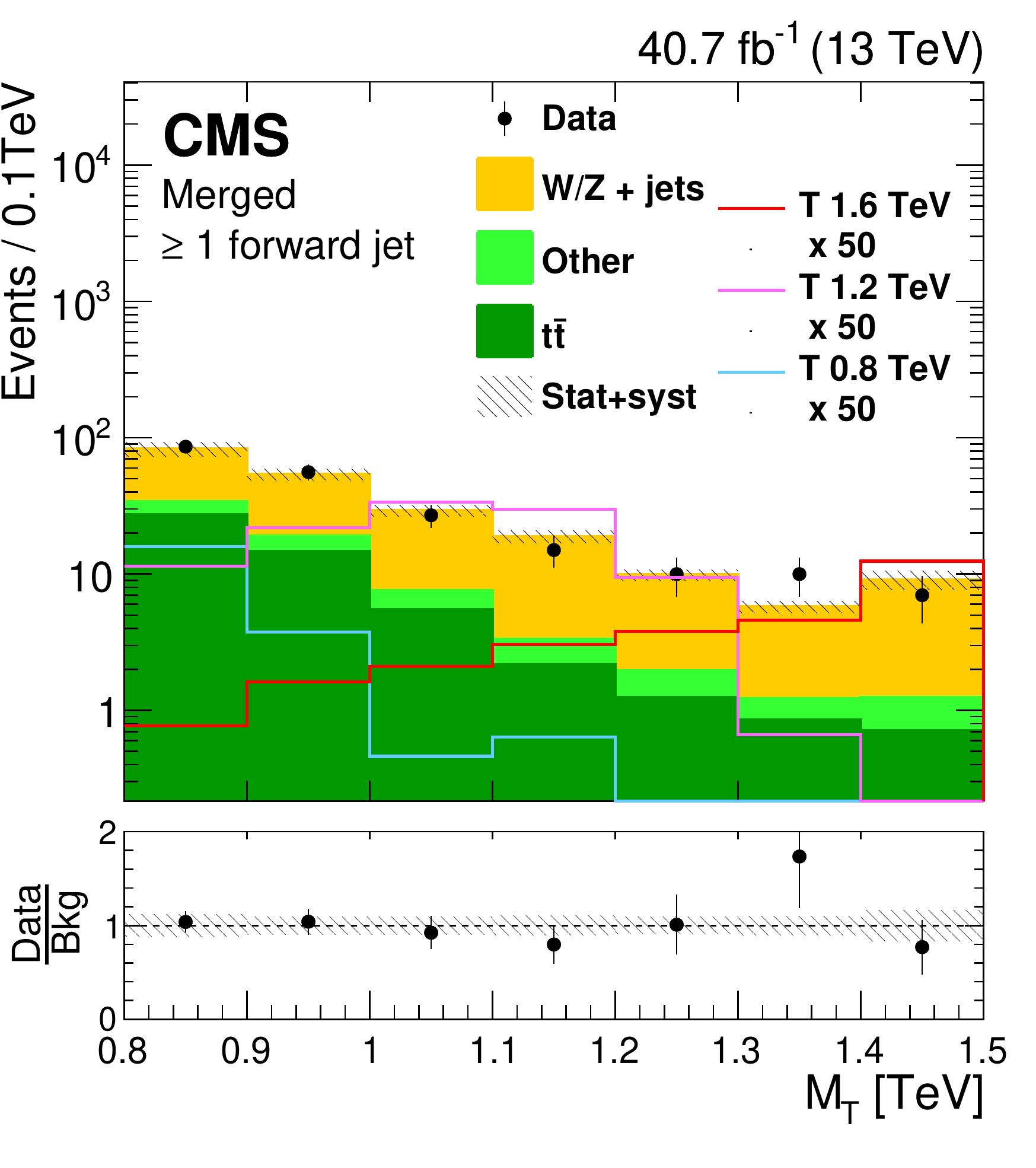}\\
\includegraphics[width=0.36\textwidth]{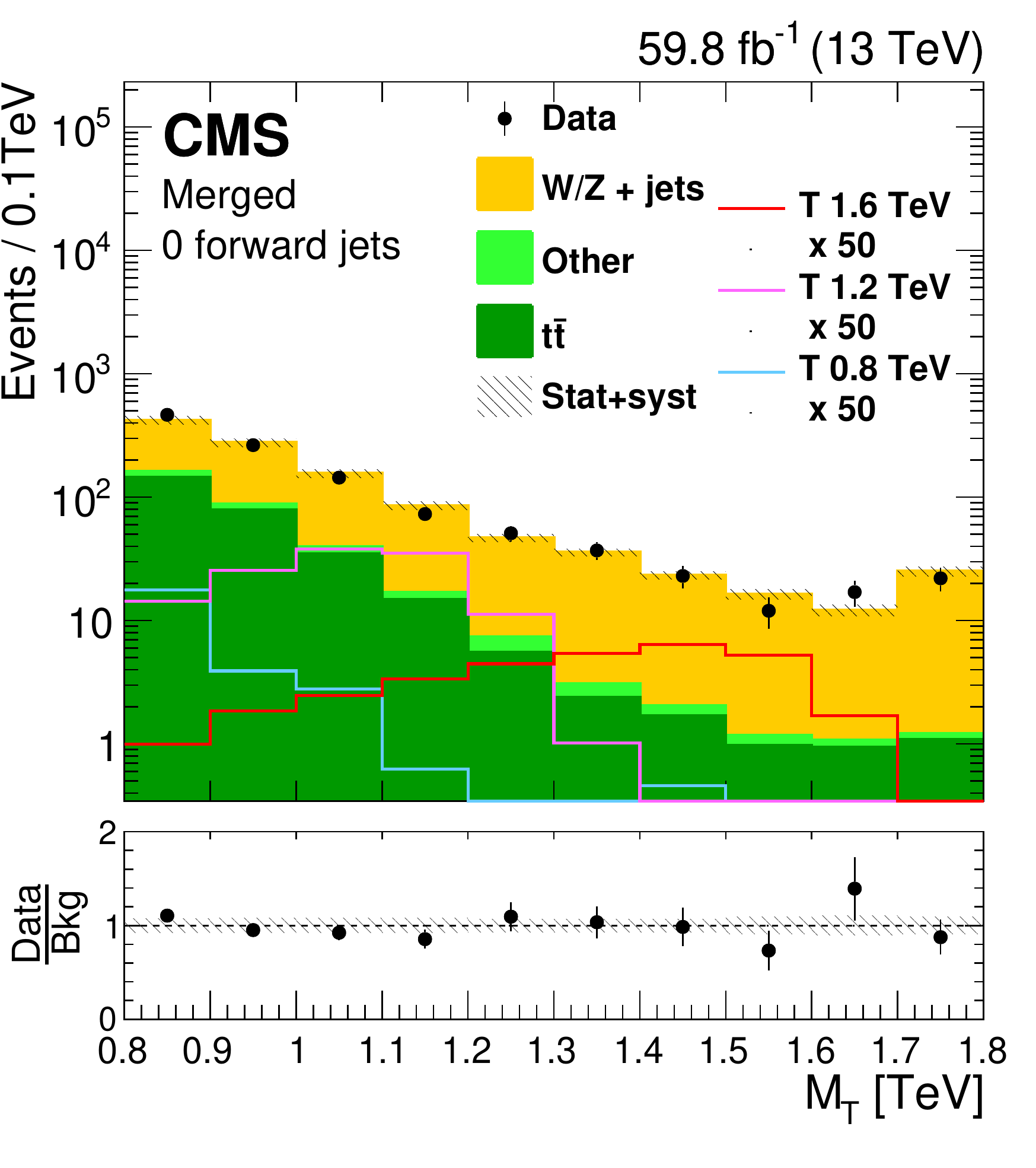}
\includegraphics[width=0.36\textwidth]{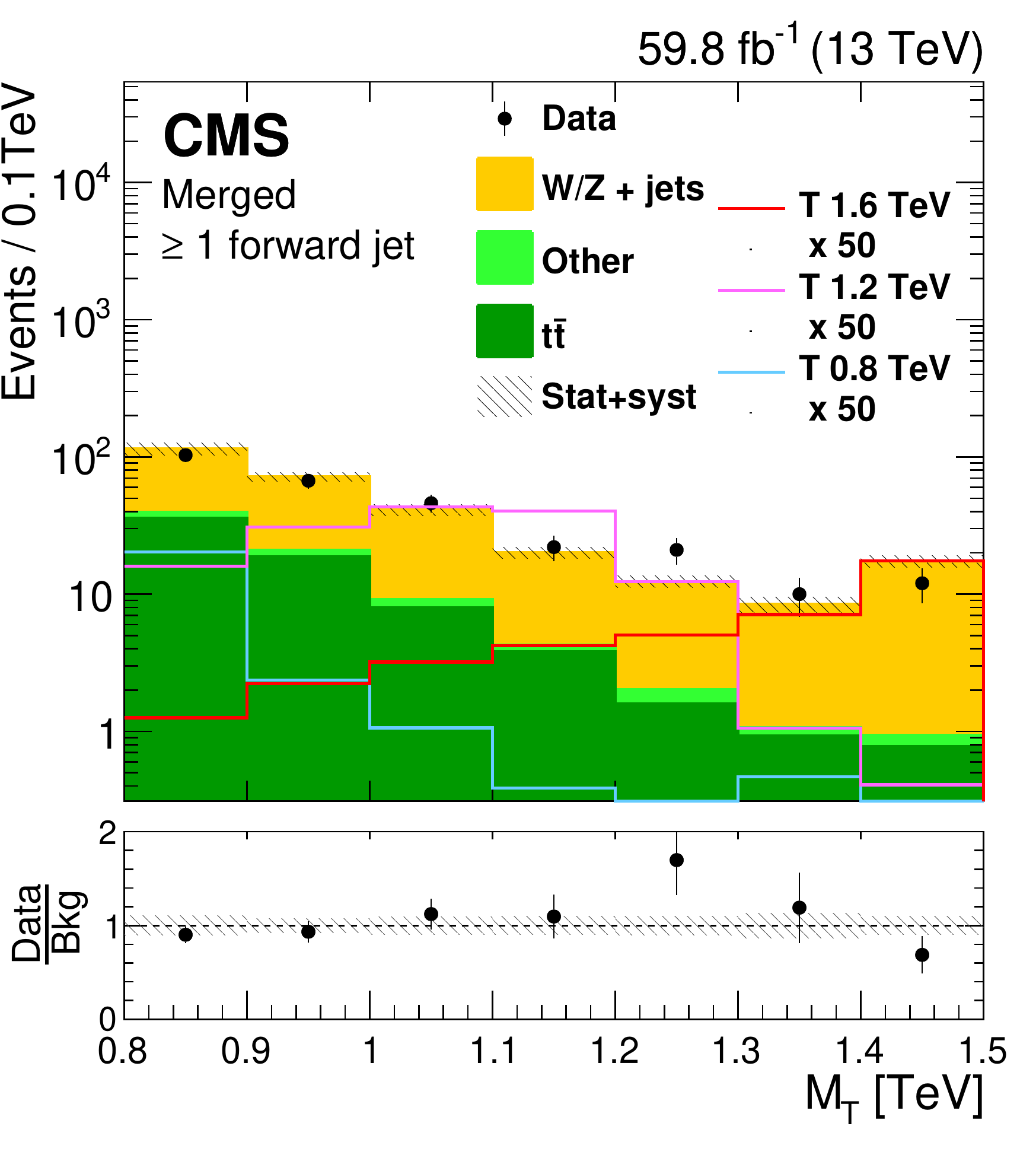}
\caption{Distributions of the transverse mass \MT of the reconstructed top quark and \ptvecmiss system, for the selected events in the merged categories,
for events with no forward jet (left) and at least one forward jet (right), and for 2016 (upper), 2017 (central), and 2018 (lower).
The overflow is included in the last bin. The distributions for the main background components have been determined in simulation with scale factors extracted from control regions. All background
processes and the respective uncertainties are derived from the fit to data, while the distributions of signal processes are represented according to the expectation before the fit. The lines show the signal predictions for three benchmark mass values (0.8, 1.2, and 1.6\TeV) of a \PQT quark of negligible resonance width. Illustrative signal yields are multiplied by a factor of 50 to improve their visibility.}
\label{fig:MtMerged}
\end{figure}

\begin{figure}[htbp!]
\centering
\includegraphics[width=0.36\textwidth]{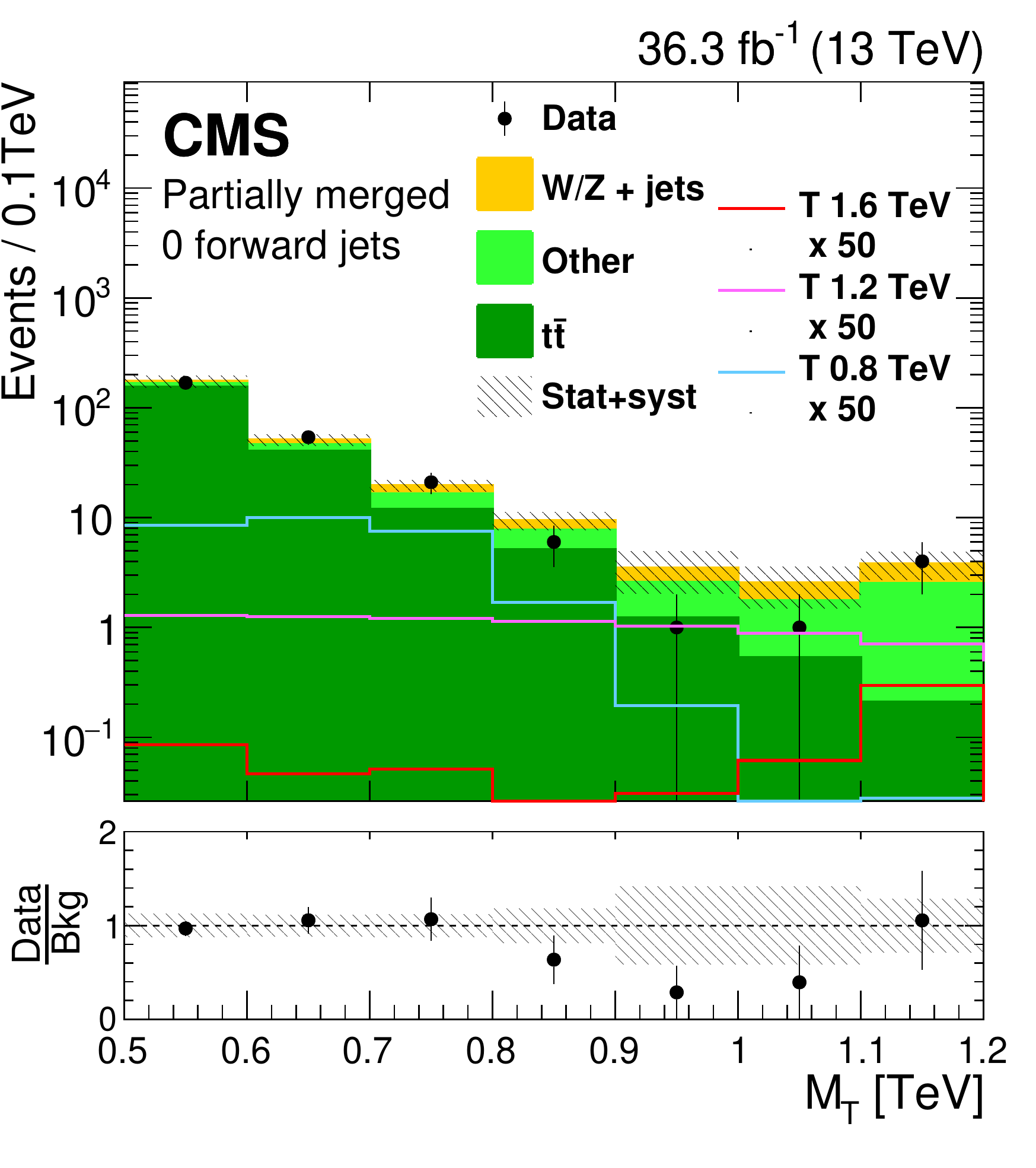}
\includegraphics[width=0.36\textwidth]{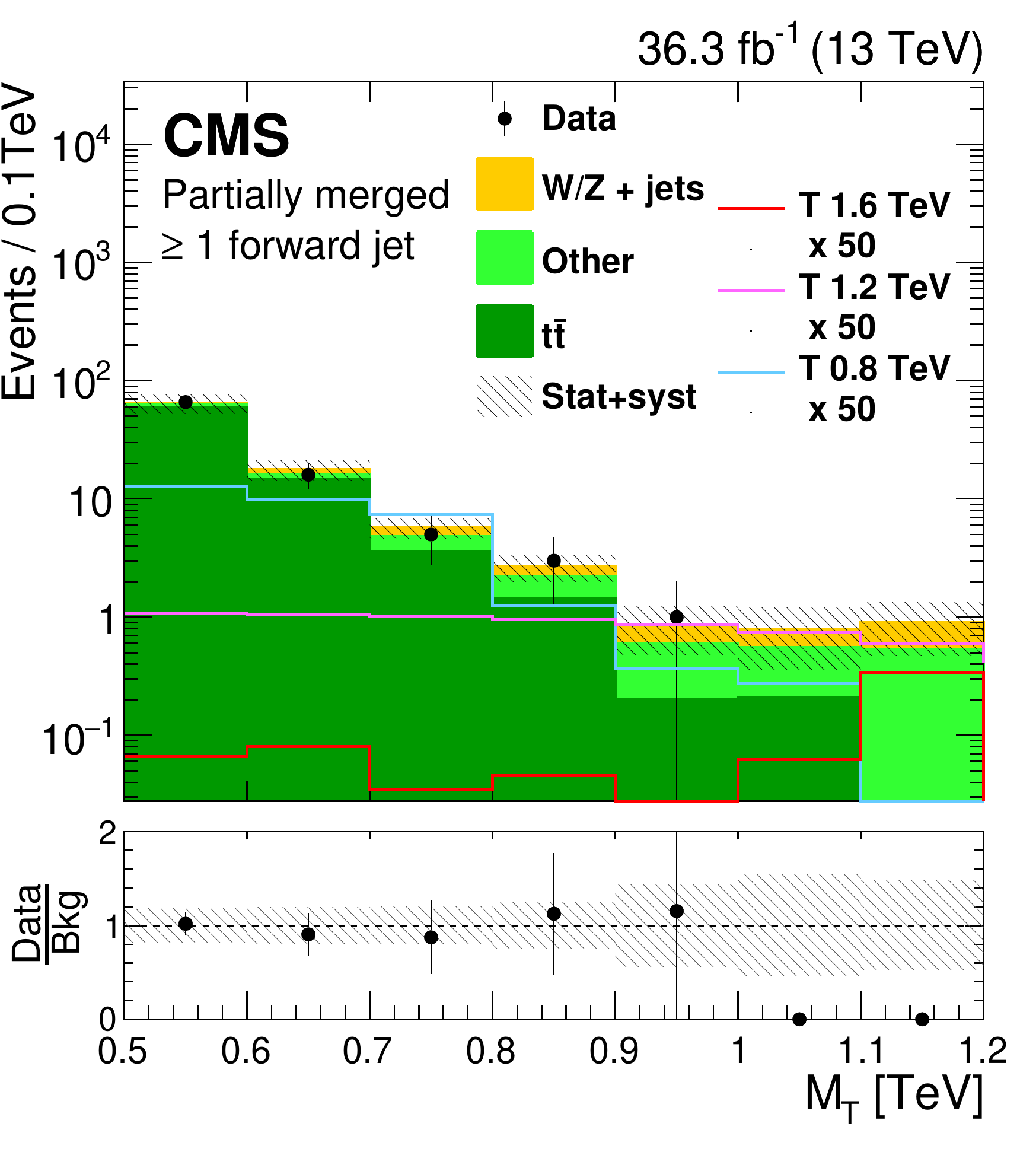}\\
\includegraphics[width=0.36\textwidth]{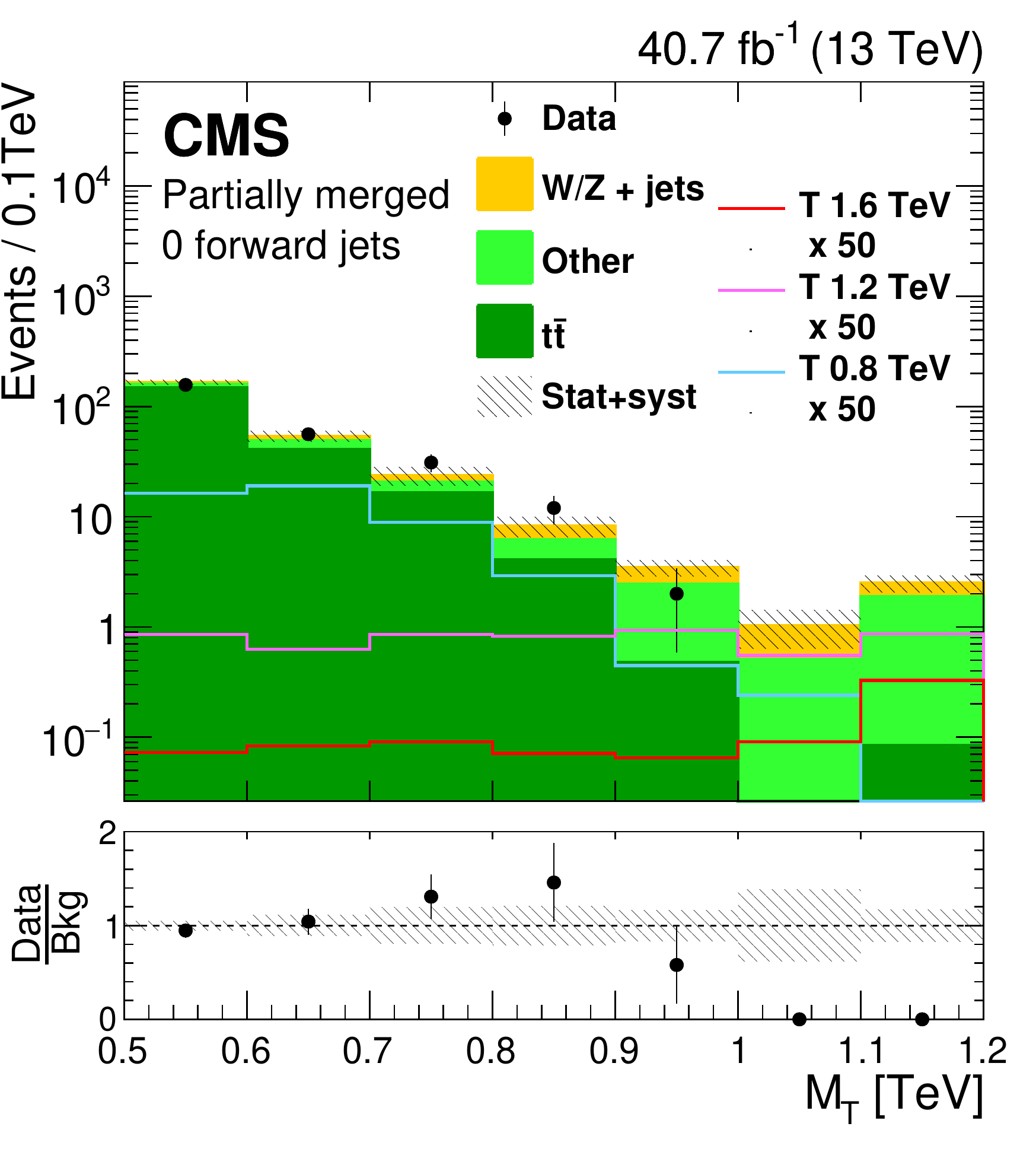}
\includegraphics[width=0.36\textwidth]{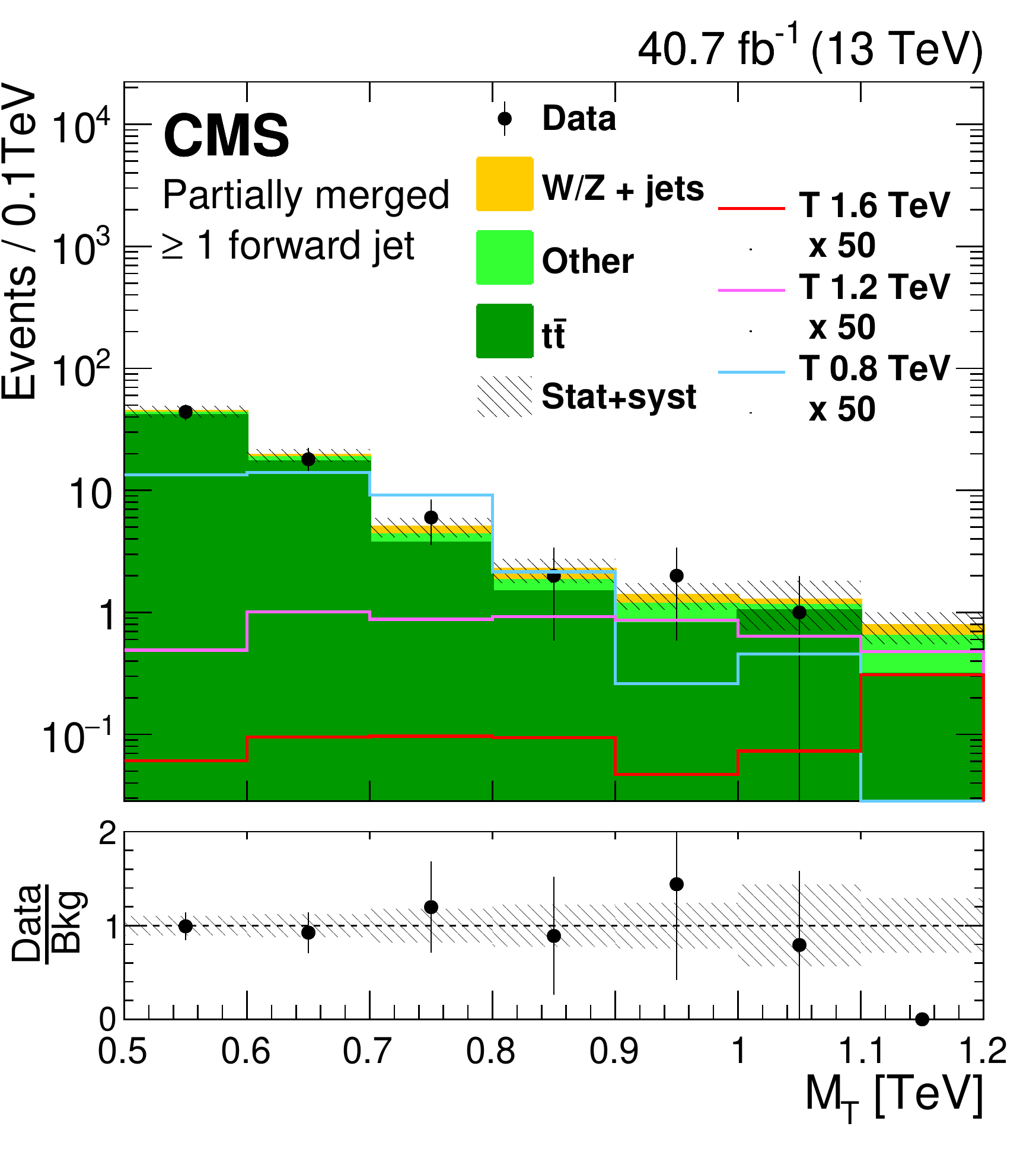}\\
\includegraphics[width=0.36\textwidth]{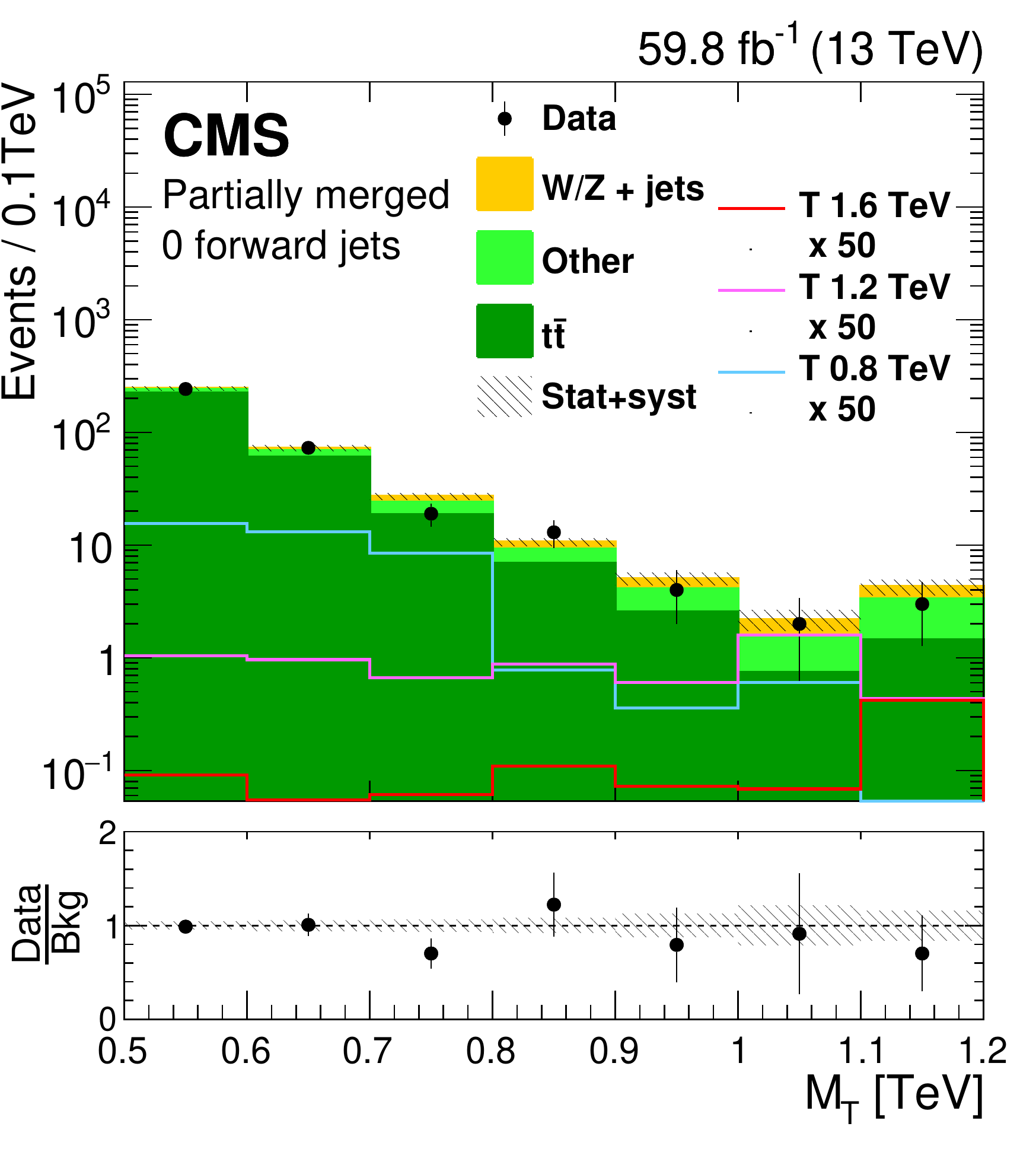}
\includegraphics[width=0.36\textwidth]{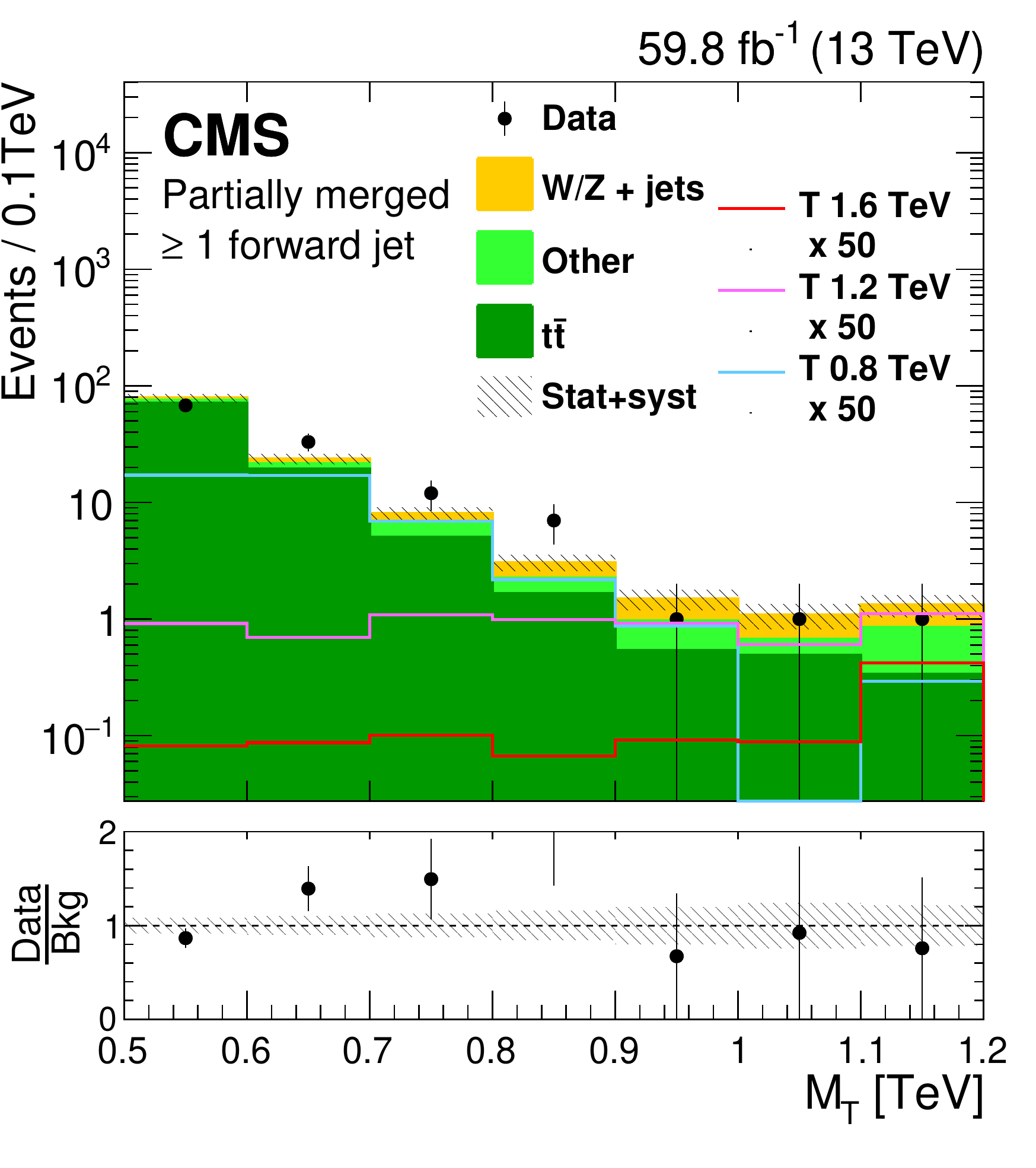}
\caption{Distributions of the transverse mass \MT of the reconstructed top quark and \ptvecmiss system, for the
 selected events in the partially merged categories, for events with no forward jet (left) and at least one forward jet (right), 
and for 2016 (upper), 2017 (central), and 2018 (lower). The overflow is included in the last bin.
The distributions for the main background components have been determined in simulation with scale factors extracted from control regions. All background
processes and the respective uncertainties are derived from the fit to data, while the distributions of signal processes are represented according to the expectation before the fit. The lines show the signal predictions for three benchmark mass values (0.8, 1.2, and 1.6\TeV) of a \PQT quark of negligible resonance width. Signal yields are multiplied by a factor of 50 to improve their visibility.
}
\label{fig:MtPartiallyMerged}
\end{figure}

\begin{figure}[htbp!]
\centering
\includegraphics[width=0.4\textwidth]{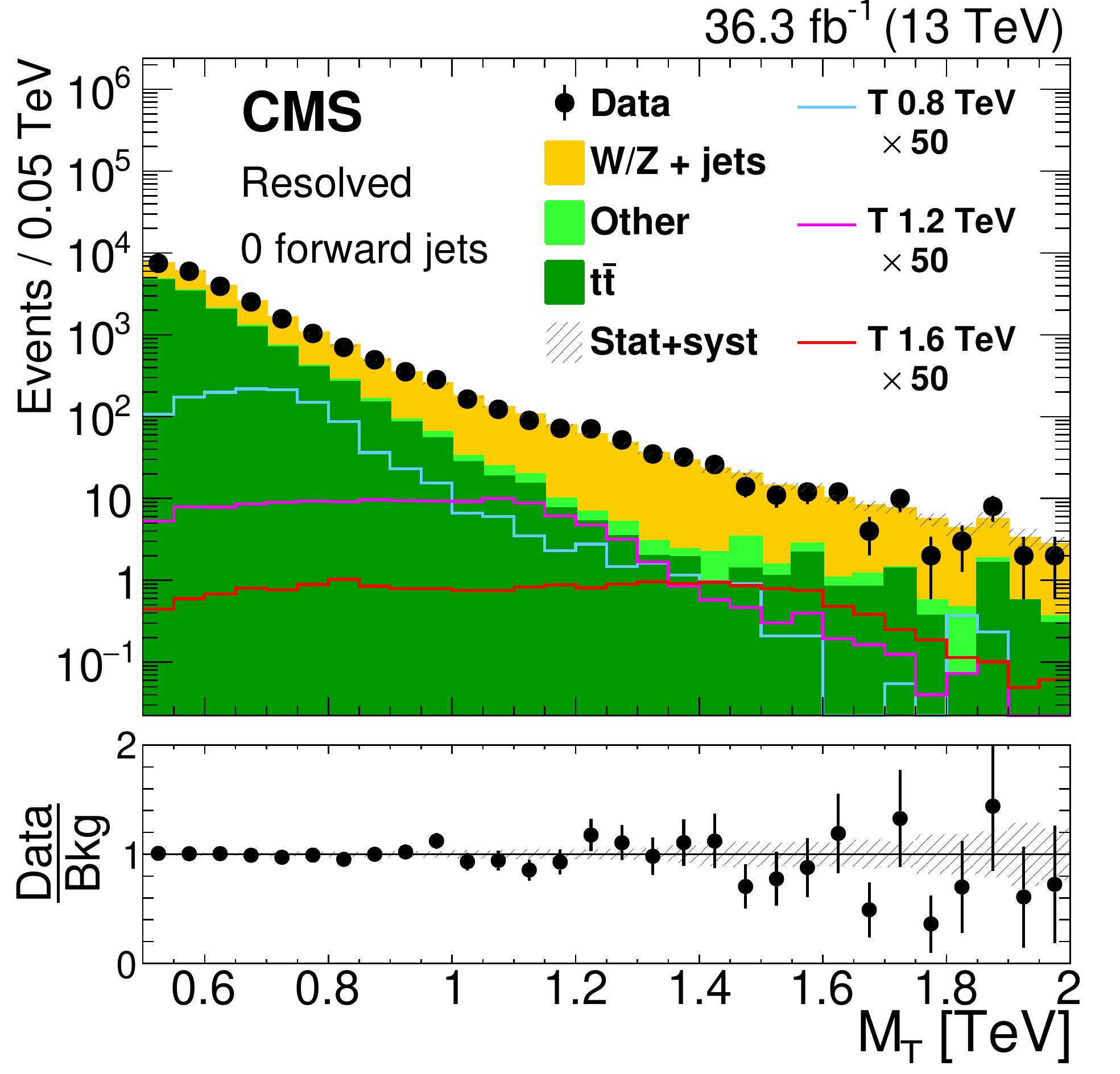}
\includegraphics[width=0.4\textwidth]{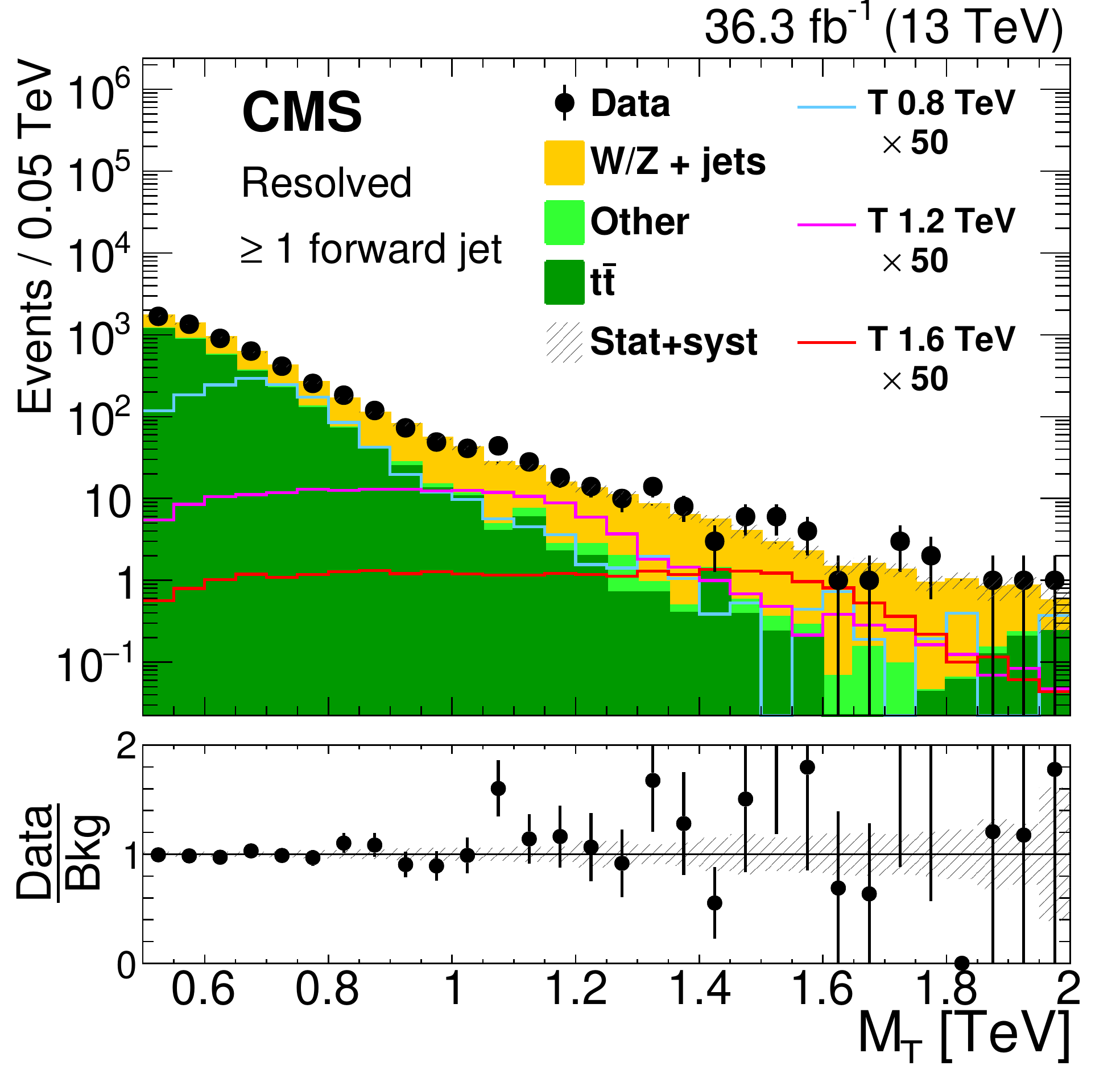}\\
\includegraphics[width=0.4\textwidth]{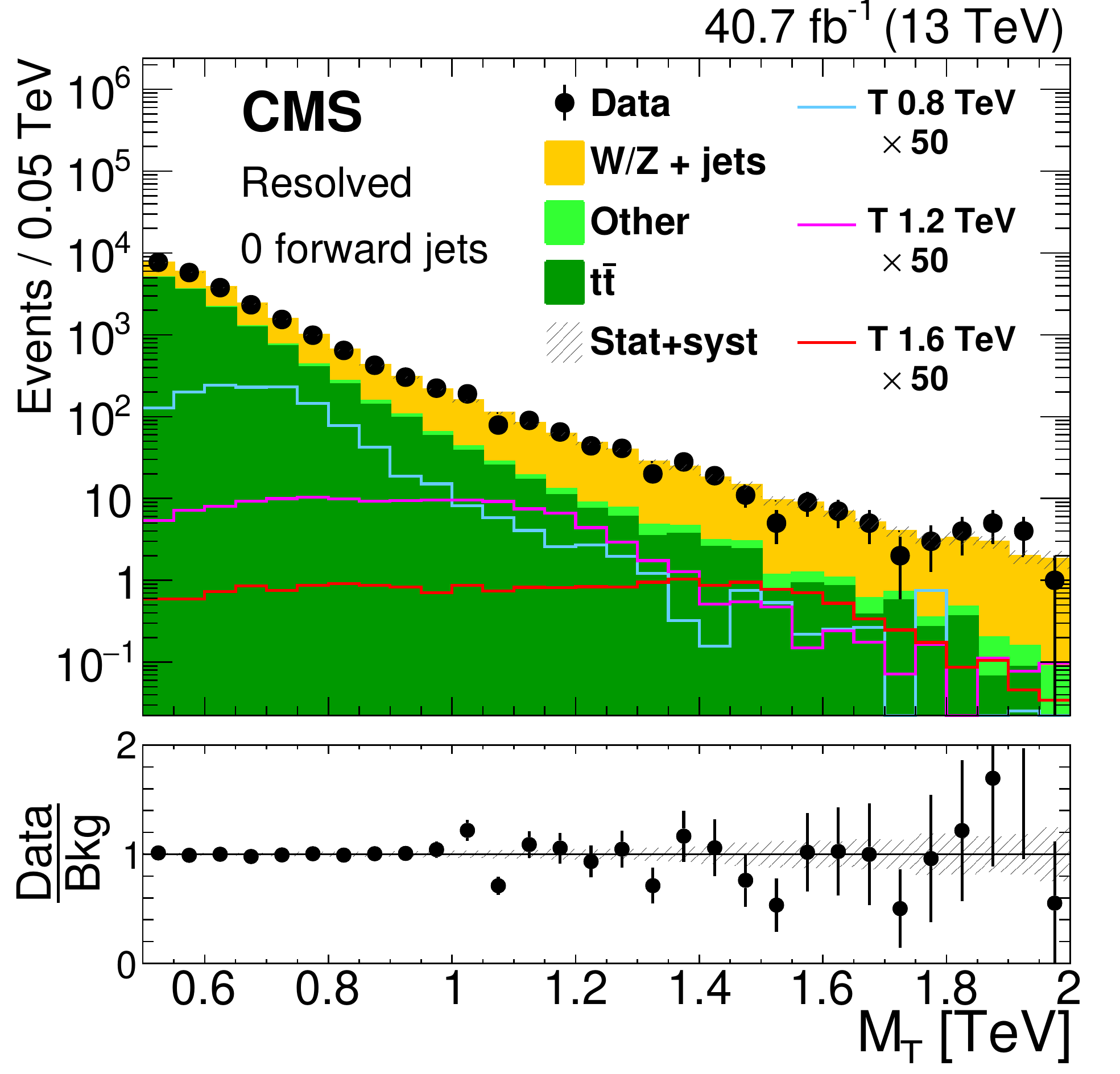}
\includegraphics[width=0.4\textwidth]{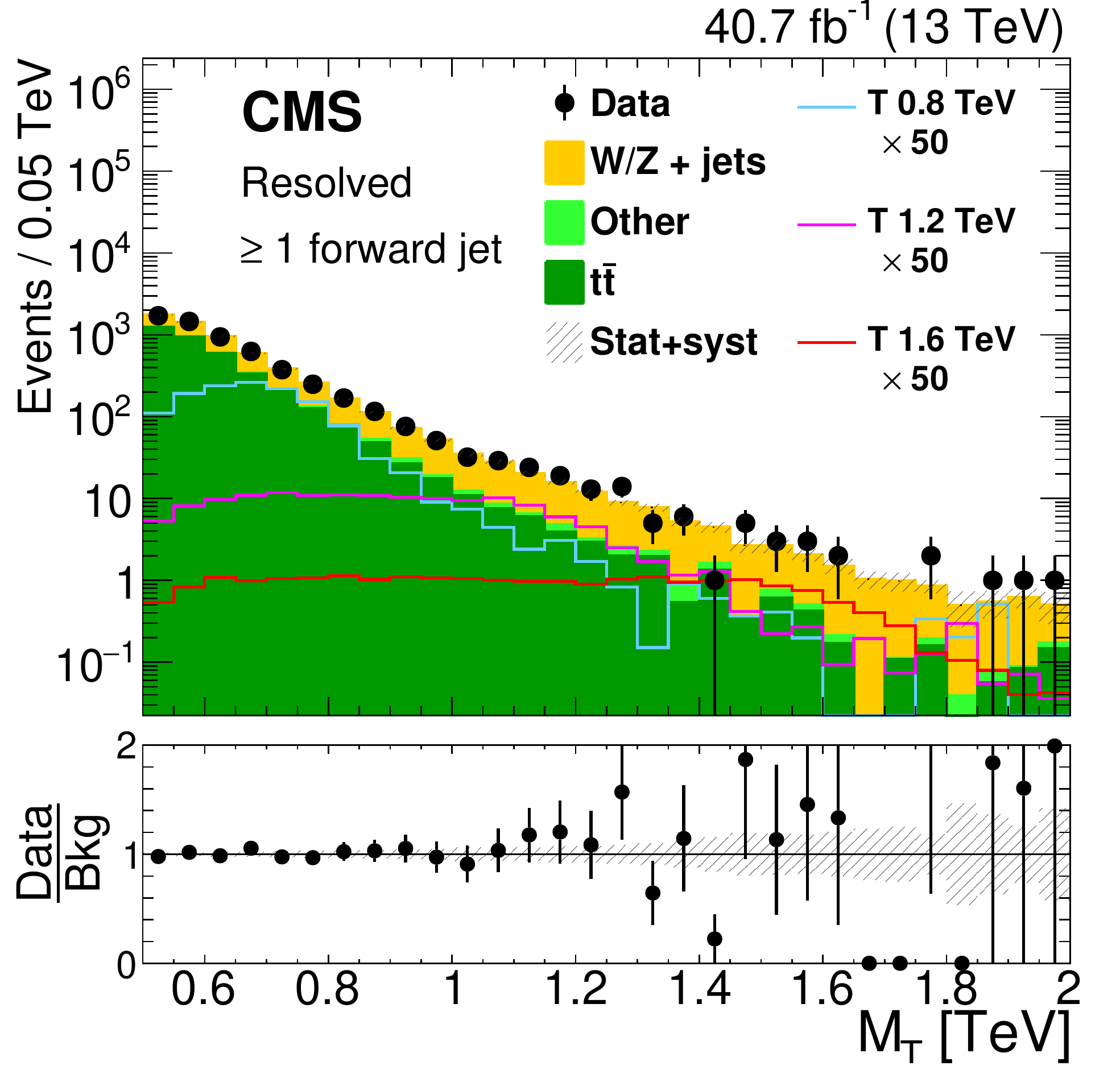}\\
\includegraphics[width=0.4\textwidth]{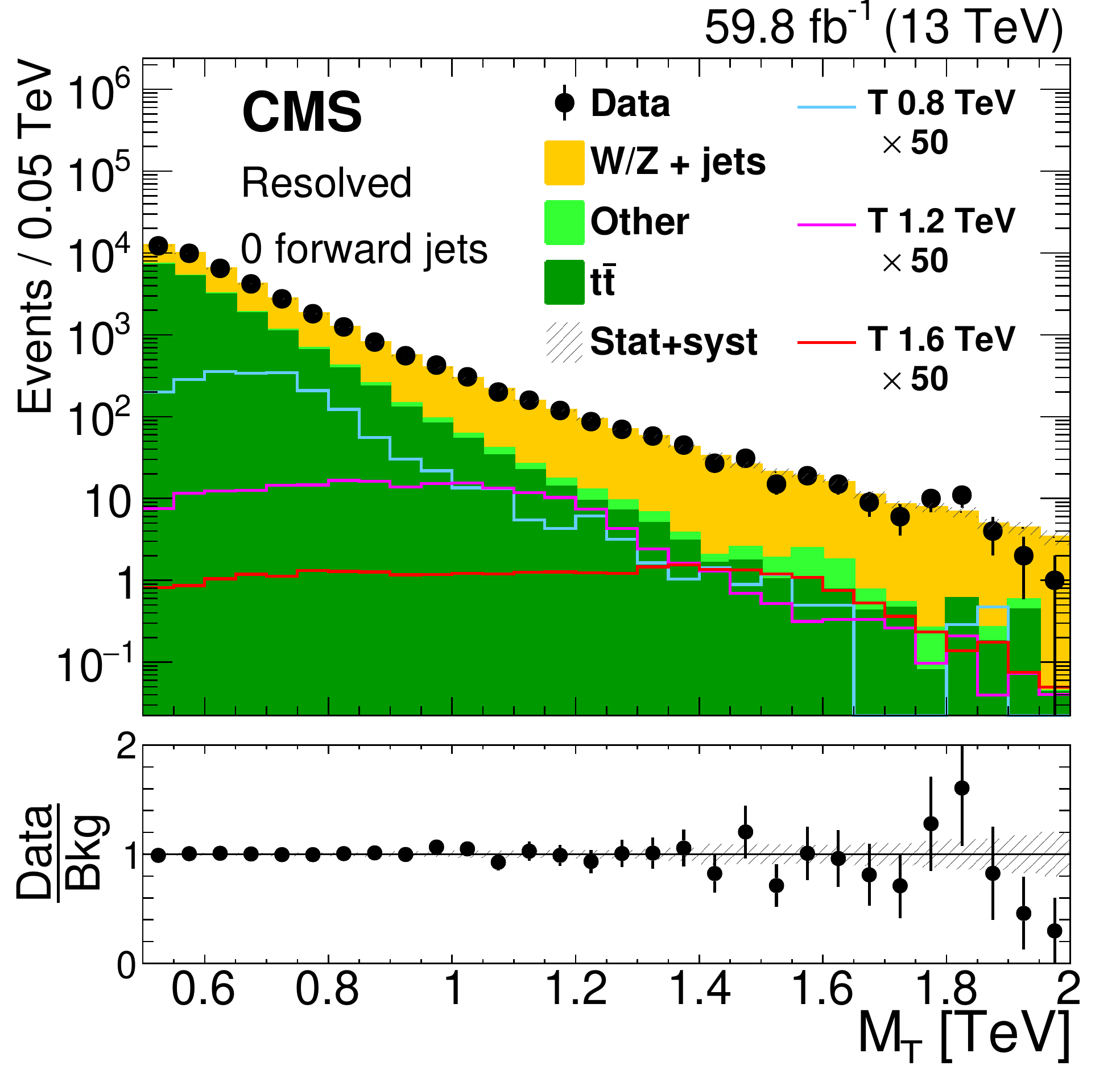}
\includegraphics[width=0.4\textwidth]{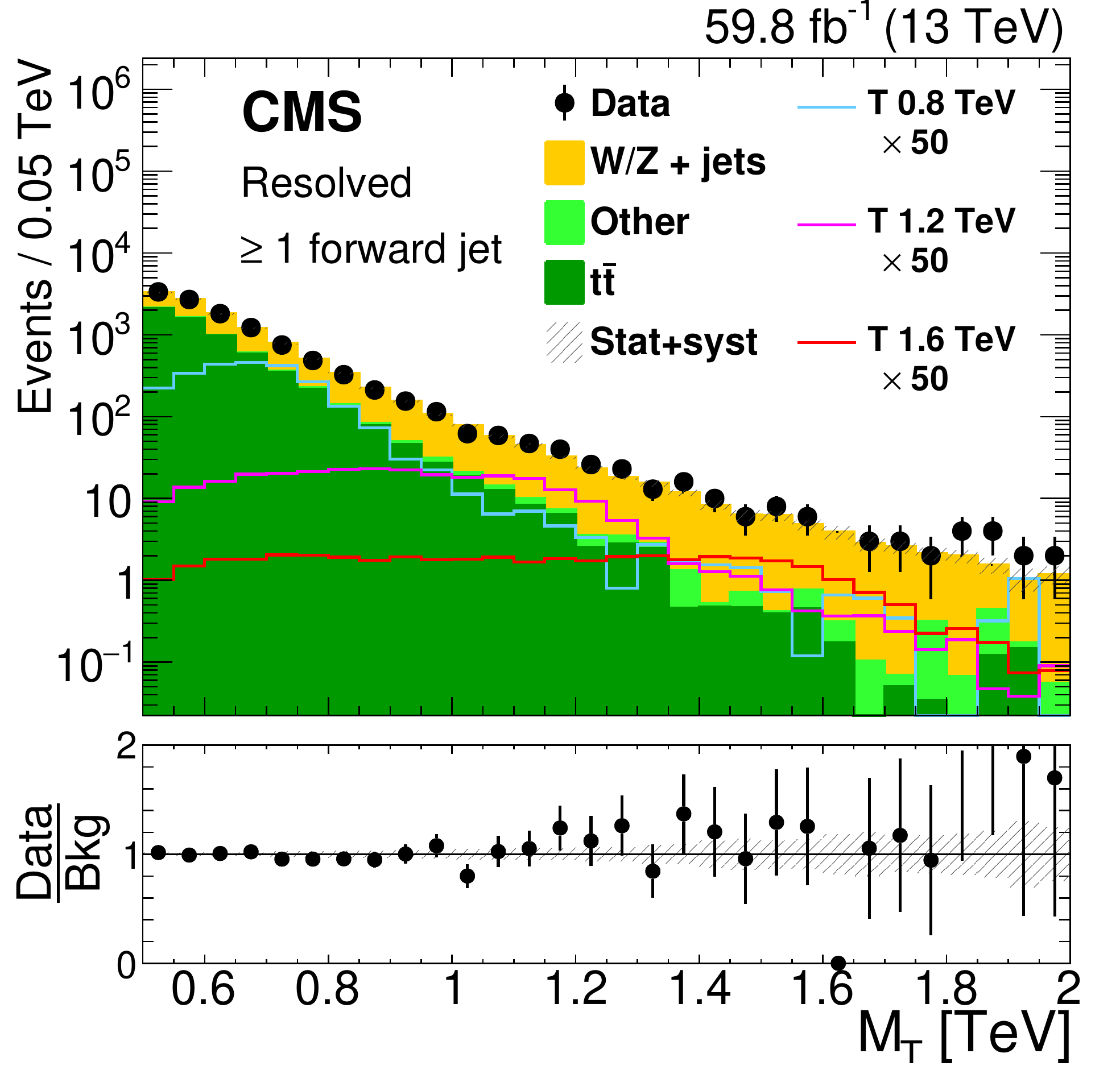}
\caption{Distributions of the transverse mass \MT of the reconstructed top quark and \ptvecmiss system, for the
selected events in the resolved categories, for events with no forward jet (left) and at least one forward jet (right), 
and for 2016 (upper), 2017 (central), and 2018 (lower).
The distributions for the main background components have been determined in simulation with scale factors extracted from control regions. All background
processes and the respective uncertainties are derived from the fit to data, while the distributions of signal processes are represented according to the expectation before the fit. The lines show the signal predictions for three benchmark mass values (0.8, 1.2, and 1.6\TeV) of a \PQT quark of negligible resonance width. Signal yields are multiplied by a factor of 50 to improve their visibility.}
\label{fig:MtResolved}
\end{figure}

The transverse mass (\MT) distributions of the selected events in the merged, partially merged, and resolved
categories are shown in Figs.~\ref{fig:MtMerged}--\ref{fig:MtResolved}, respectively. The predictions for the background processes have been obtained with the procedure described in Section~\ref{sec:background}.

The signal extraction procedure is based on a simultaneous maximum likelihood fit
to the \MT distributions in all signal regions.
Systematic uncertainties are treated as nuisance parameters, assumed to be described by a log-normal
probability density function, for systematic uncertainties that affect the rate, or a Gaussian
probability density function, for systematic uncertainties that affect the \MT distribution shape.
Templates for background and signal are also shown in Figs.~\ref{fig:MtMerged}--\ref{fig:MtResolved}.

No evidence of a statistically significant signal is found from the fit to the data.
The largest excess over the expected background, corresponding to a
significance of 2.5 standard deviations, is found in the 
narrow-width hypothesis for a \PQT quark mass of 1.4\TeV.
The global significance to have such an excess over the entire mass range 0.6--1.8\TeV is 2.2 standard deviations.
This deviation from the expected background is mainly due to a broad excess of events
in the transverse mass distribution for $\MT > 1\TeV $, observed in 2016 data 
for events in the resolved category and with at least one forward jet (see Fig.~\ref{fig:MtResolved}).

Upper limits at 95\% confidence level (\CL) are set
on the product of the single \PQT quark production cross section and
$\PQT \to \tZ$ channel branching fraction. The limits are determined using the
\CLs method~\cite{Junk:1999kv,Read:2002hq,Cowan:2010js,ATLAS:2011tau},
taking the profile likelihood ratio as the test statistic with the asymptotic limit approximation.

\begin{figure}[htbp!]
    \centering
    \includegraphics[width=0.49\textwidth]{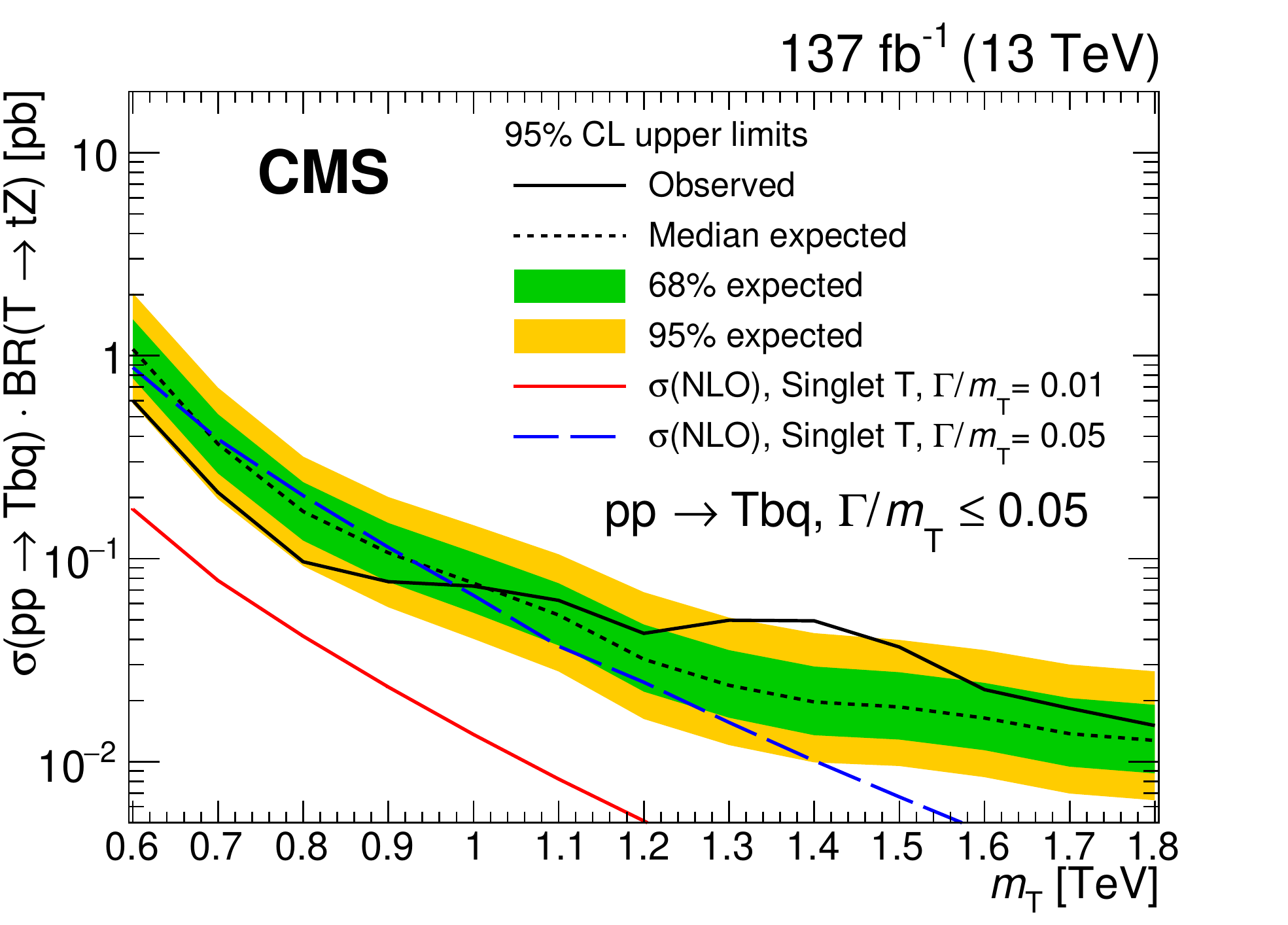}
    \includegraphics[width=0.49\textwidth]{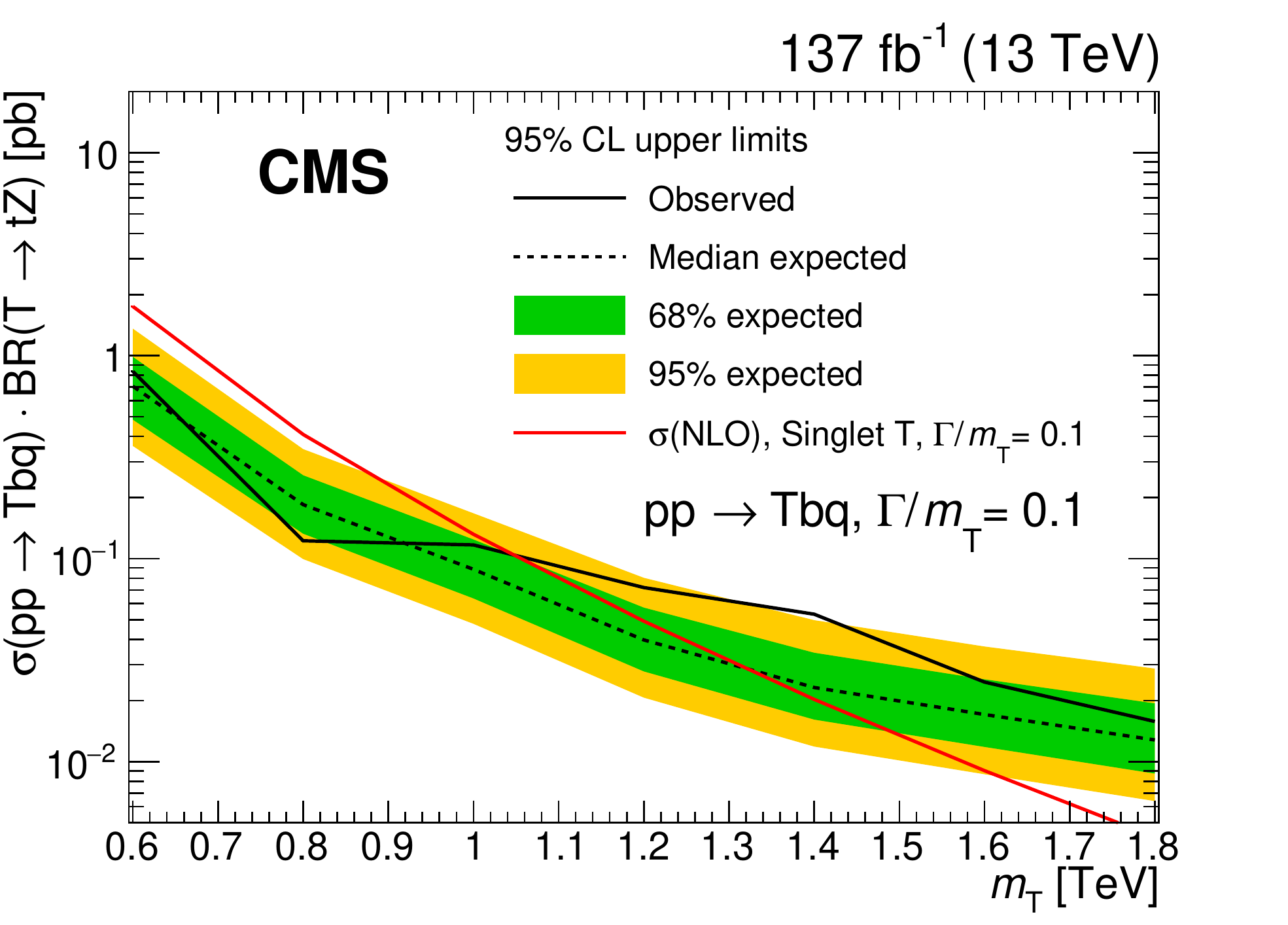}
    \includegraphics[width=0.49\textwidth]{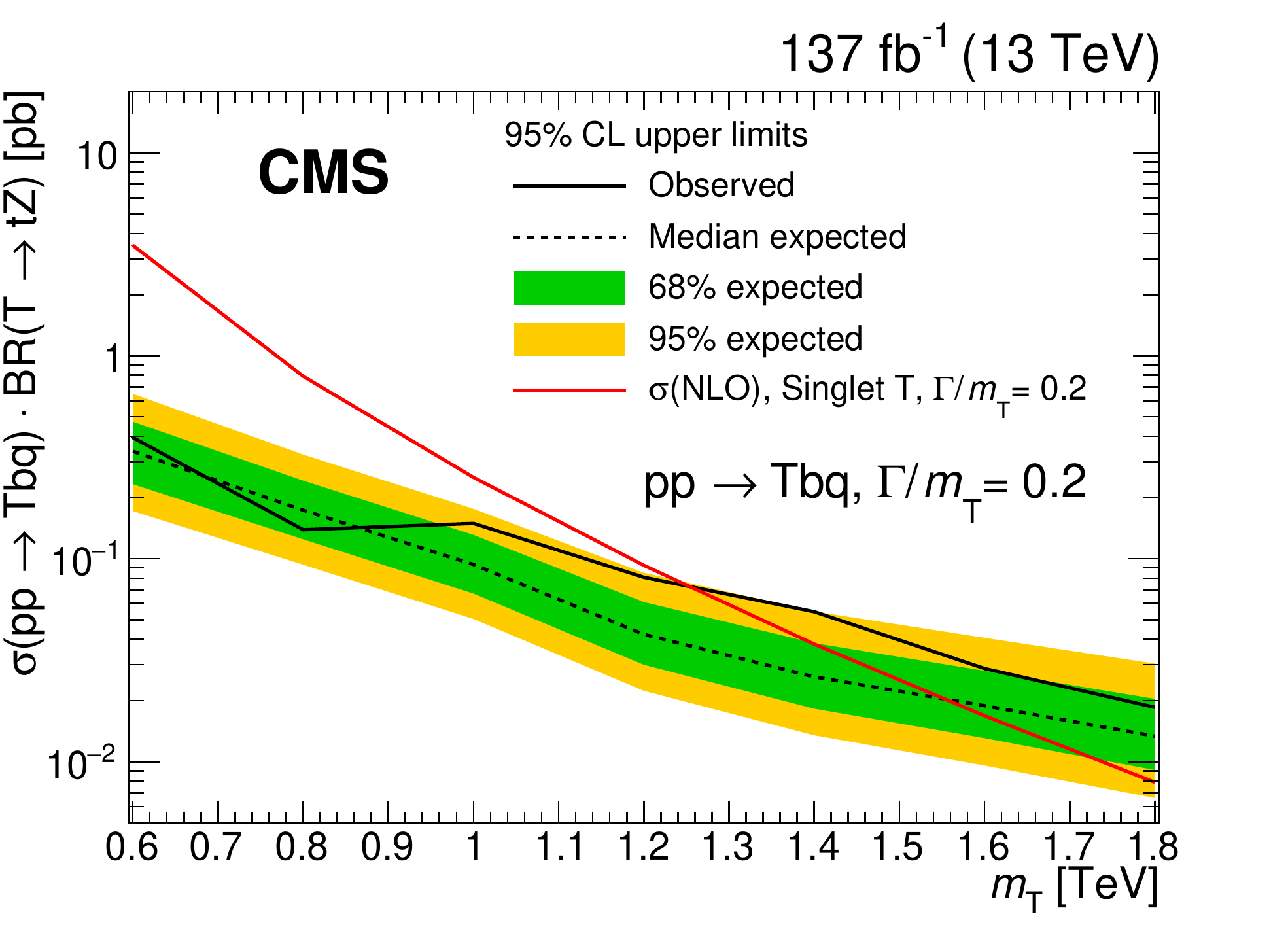}
    \includegraphics[width=0.49\textwidth]{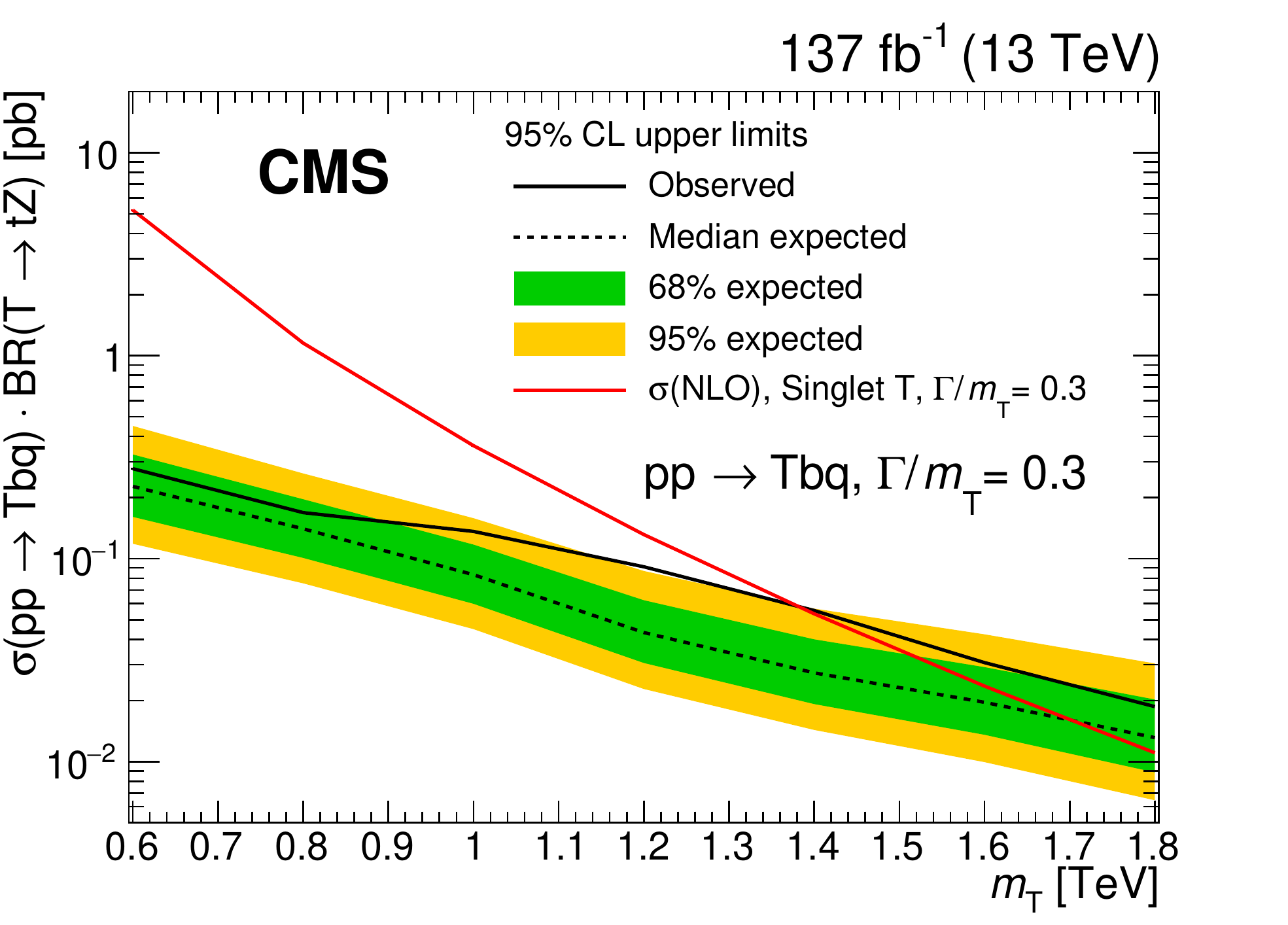}
    \caption{Observed and expected 95\% \CL upper limits on the product of the 
    single production cross section for a singlet VLQ \PQT quark and the $\PQT \to \tZ$ 
    branching fraction, as functions of the \PQT quark mass $m_{\PQT}$ for a narrow-width
    resonance (upper left), and a width of 10\% (upper right), 20\% (lower left), 
    and 30\% (lower right) of the \PQT quark mass. A singlet \PQT quark is assumed,
    produced in association with a bottom quark.
    The inner (green) band and the outer (yellow) band indicate the regions
    containing 68 and 95\%, respectively, of the distribution of
    limits expected under the background-only hypothesis.
    The solid curves show the theoretical expectation at NLO.
    In the case of a narrow-width resonance, the width of 1\,(5)\% 
    of the resonance mass is indicated with a red (blue) curve.}
    \label{fig:combinedUL}
    \end{figure}

\begin{figure}[htbp!]
    \centering
    \includegraphics[width=0.72\textwidth]{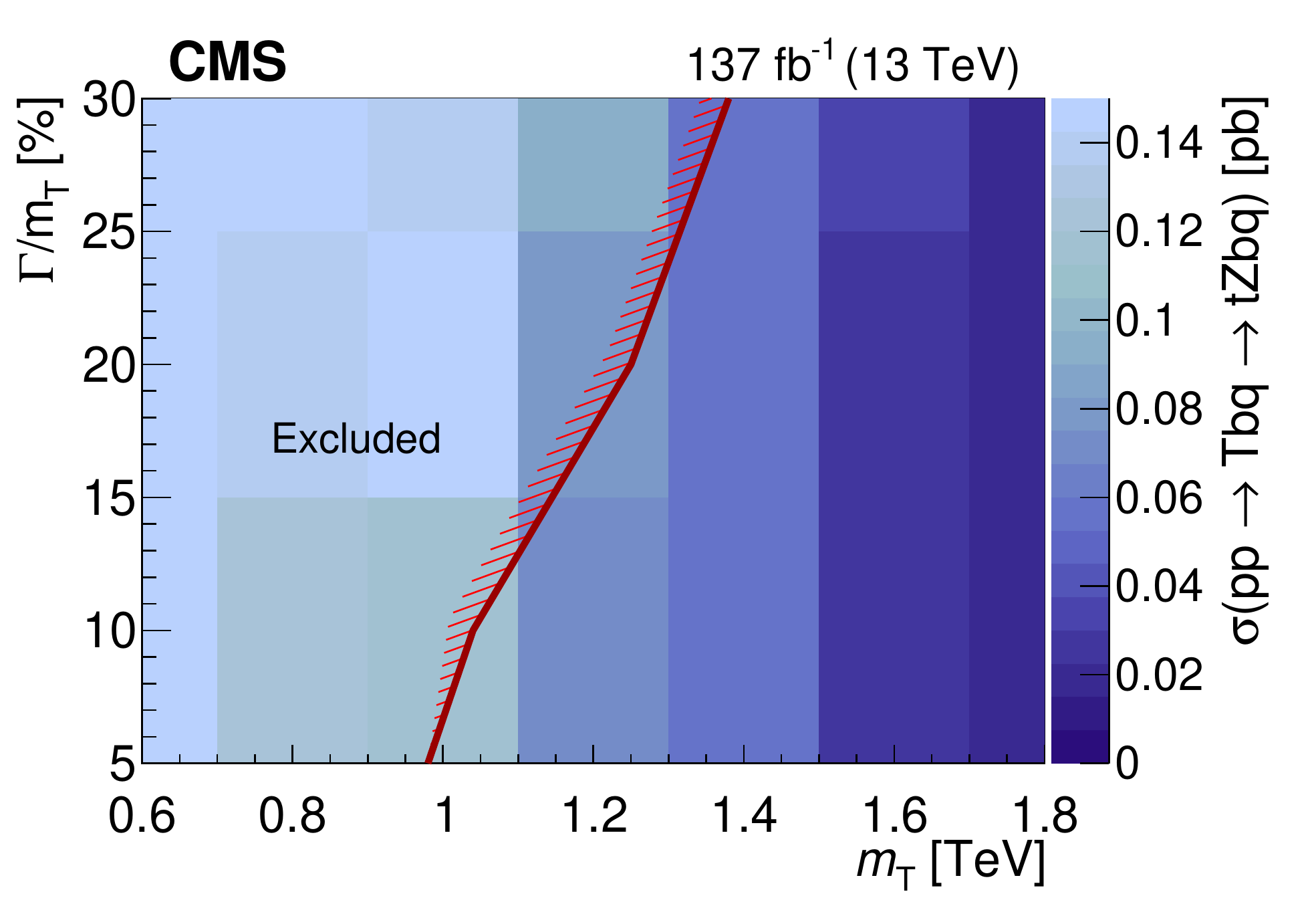}
    \caption{Observed 95\% \CL upper limit on the product of the single production cross section for a singlet VLQ \PQT quark and the $\PQT \to \PQt \PZ$ branching fraction, as a function of the \PQT quark mass $m_{\PQT}$ and width $\Gamma$, for widths from 5
    to 30\% of the mass. A singlet \PQT quark that is produced in association with a bottom quark is assumed.
    The solid red line indicates the boundary of the excluded
    region (on the hatched side) of theoretical cross sections,
    as reported in Table~\ref{tab:CrossSec}.}
    \label{fig:combinedUL2D}
    \end{figure}

The observed and expected combined upper limits from the six event categories,
as functions of the \PQT quark mass $m_{\PQT}$, and for several hypotheses on its width,
are shown in Figs.~\ref{fig:combinedUL} and ~\ref{fig:combinedUL2D} respectively.
The limits at low values of $m_{\PQT}$ are mostly driven by the resolved categories, 
while the merged categories are the most effective at high values, reflecting
the different sensitivities in different $m_{\PQT}$ ranges.
The limits at intermediate values of $m_{\PQT}$ benefit from the sensitivity 
of all the event categories employed in the analysis. The high limit at 1.4\TeV is
driven by the excess seen in the resolved category.

Assuming a narrow-width resonance, values of the product of production cross section and branching fraction greater than 602 to 15\unit{fb} are excluded at 95\% \CL
for masses in the range 0.6--1.8\TeV.

For a resonance of width in the range 10--30\% of its mass, 
values greater than 836 to 16\unit{fb} are excluded at 95\% \CL
for masses in the range 0.6--1.8\TeV.

These results set the most stringent limits on the product of the cross section for
single \PQT quark production and the branching fraction into the $\PQt\PZ$ decay channel
over the mass ranges 0.6--1.2 and 1.5--1.8\TeV.
For mass values in the range 1.2--1.5\TeV, the measurements reported
in~\cite{Sirunyan:2019xeh,Sirunyan:2017ynj,Aaboud:2018zpr} yield give a comparable
or lower limit because of the excess of events observed in this region in our data.

These results also set the best lower limits on the \PQT quark mass
in the singlet model, for various resonance width hypotheses:
values of \PQT quark mass lower than 0.98, 1.1, 1.3,
and 1.4\TeV are excluded for resonance widths 5, 10, 20, and 30\% of the
mass, respectively.

\section{Summary}\label{sec:summary}
A search for the single production of a vector-like quark \PQT 
with charge~$2/3$\,$e$ decaying to a top quark and a \PZ boson has been presented.
The analysis is based on LHC proton-proton collision data collected by the CMS experiment, 
corresponding to an integrated luminosity of 137\fbinv.
Upper limits at 95\% confidence level are set
on the product of the production cross section and the
$\PQT \to \tZ$ channel branching fraction. Values greater
than 602 to 15\unit{fb} for \PQT quark masses
between 0.6 and 1.8\TeV are excluded at 95\% confidence level for a
\PQT quark of negligible resonance width produced in association with a bottom quark.
Values greater than 836 to 16\unit{fb} for masses between 0.6 and 1.8\TeV
are excluded at 95\% confidence level for
a \PQT quark of resonance width from 10 to 30\% of its mass.
These results provide the best exclusion limits on the production of single
vector-like \PQT quarks in the $\tZ$ decay channel over the mass
range from 0.6 to 1.2\TeV and from 1.5 to 1.8\TeV. 
An interpretation of these results within a theoretical framework 
in which the \PQT quark is a singlet, and assuming a resonance width of 5\% 
of the mass, leads to the exclusion of a \PQT quark of mass below 0.98\TeV. 
The excluded mass range extends up to 1.4\TeV for a resonance width 30\%
of the mass. This is the first search for single \PQT quark production
based on the full LHC Run 2 data set.

\begin{acknowledgments}
We congratulate our colleagues in the CERN accelerator departments for the excellent performance of the LHC and thank the technical and administrative staffs at CERN and at other CMS institutes for their contributions to the success of the CMS effort. In addition, we gratefully acknowledge the computing cemters and personnel of the Worldwide LHC Computing Grid and other cemters for delivering so effectively the computing infrastructure essential to our analyses. Finally, we acknowledge the enduring support for the construction and operation of the LHC, the CMS detector, and the supporting computing infrastructure provided by the following funding agencies: BMBWF and FWF (Austria); FNRS and FWO (Belgium); CNPq, CAPES, FAPERJ, FAPERGS, and FAPESP (Brazil); MES and BNSF (Bulgaria); CERN; CAS, MoST, and NSFC (China); MINCIENCIAS (Colombia); MSES and CSF (Croatia); RIF (Cyprus); SENESCYT (Ecuador); MoER, ERC PUT and ERDF (Estonia); Academy of Finland, MEC, and HIP (Finland); CEA and CNRS/IN2P3 (France); BMBF, DFG, and HGF (Germany); GSRI (Greece); NKFIA (Hungary); DAE and DST (India); IPM (Iran); SFI (Ireland); INFN (Italy); MSIP and NRF (Republic of Korea); MES (Latvia); LAS (Lithuania); MOE and UM (Malaysia); BUAP, CINVESTAV, CONACYT, LNS, SEP, and UASLP-FAI (Mexico); MOS (Montenegro); MBIE (New Zealand); PAEC (Pakistan); MSHE and NSC (Poland); FCT (Portugal); JINR (Dubna); MON, RosAtom, RAS, RFBR, and NRC KI (Russia); MESTD (Serbia); MCIN/AEI and PCTI (Spain); MOSTR (Sri Lanka); Swiss Funding Agencies (Switzerland); MST (Taipei); ThEPCenter, IPST, STAR, and NSTDA (Thailand); TUBITAK and TAEK (Turkey); NASU (Ukraine); STFC (United Kingdom); DOE and NSF (USA).

\hyphenation{Rachada-pisek} Individuals have received support from the Marie-Curie program and the European Research Council and Horizon 2020 Grant, contract Nos.\ 675440, 724704, 752730, 758316, 765710, 824093, 884104, and COST Action CA16108 (European Union); the Leventis Foundation; the Alfred P.\ Sloan Foundation; the Alexander von Humboldt Foundation; the Belgian Federal Science Policy Office; the Fonds pour la Formation \`a la Recherche dans l'Industrie et dans l'Agriculture (FRIA-Belgium); the Agentschap voor Innovatie door Wetenschap en Technologie (IWT-Belgium); the F.R.S.-FNRS and FWO (Belgium) under the ``Excellence of Science -- EOS" -- be.h project n.\ 30820817; the Beijing Municipal Science \& Technology Commission, No. Z191100007219010; the Ministry of Education, Youth and Sports (MEYS) of the Czech Republic; the Deutsche Forschungsgemeinschaft (DFG), under Germany's Excellence Strategy -- EXC 2121 ``Quantum Universe" -- 390833306, and under project number 400140256 - GRK2497; the Lend\"ulet (``Momentum") Program and the J\'anos Bolyai Research Scholarship of the Hungarian Academy of Sciences, the New National Excellence Program \'UNKP, the NKFIA research grants 123842, 123959, 124845, 124850, 125105, 128713, 128786, and 129058 (Hungary); the Council of Science and Industrial Research, India; the Latvian Council of Science; the Ministry of Science and Higher Education and the National Science Center, contracts Opus 2014/15/B/ST2/03998 and 2015/19/B/ST2/02861 (Poland); the Funda\c{c}\~ao para a Ci\^encia e a Tecnologia, grant CEECIND/01334/2018 (Portugal); the National Priorities Research Program by Qatar National Research Fund; the Ministry of Science and Higher Education, projects no. 14.W03.31.0026 and no. FSWW-2020-0008, and the Russian Foundation for Basic Research, project No.19-42-703014 (Russia); MCIN/AEI/10.13039/501100011033, ERDF ``a way of making Europe", and the Programa Estatal de Fomento de la Investigaci{\'o}n Cient{\'i}fica y T{\'e}cnica de Excelencia Mar\'{\i}a de Maeztu, grant MDM-2017-0765 and Programa Severo Ochoa del Principado de Asturias (Spain); the Stavros Niarchos Foundation (Greece); the Rachadapisek Sompot Fund for Postdoctoral Fellowship, Chulalongkorn University and the Chulalongkorn Academic into Its 2nd Century Project Advancement Project (Thailand); the Kavli Foundation; the Nvidia Corporation; the SuperMicro Corporation; the Welch Foundation, contract C-1845; and the Weston Havens Foundation (USA).

\end{acknowledgments}

\bibliography{auto_generated}

\cleardoublepage \appendix\section{The CMS Collaboration \label{app:collab}}\begin{sloppypar}\hyphenpenalty=5000\widowpenalty=500\clubpenalty=5000\cmsinstitute{Yerevan~Physics~Institute, Yerevan, Armenia}
A.~Tumasyan
\cmsinstitute{Institut~f\"{u}r~Hochenergiephysik, Vienna, Austria}
W.~Adam\cmsorcid{0000-0001-9099-4341}, J.W.~Andrejkovic, T.~Bergauer\cmsorcid{0000-0002-5786-0293}, S.~Chatterjee\cmsorcid{0000-0003-2660-0349}, M.~Dragicevic\cmsorcid{0000-0003-1967-6783}, A.~Escalante~Del~Valle\cmsorcid{0000-0002-9702-6359}, R.~Fr\"{u}hwirth\cmsAuthorMark{1}, M.~Jeitler\cmsAuthorMark{1}\cmsorcid{0000-0002-5141-9560}, N.~Krammer, L.~Lechner\cmsorcid{0000-0002-3065-1141}, D.~Liko, I.~Mikulec, P.~Paulitsch, F.M.~Pitters, J.~Schieck\cmsAuthorMark{1}\cmsorcid{0000-0002-1058-8093}, R.~Sch\"{o}fbeck\cmsorcid{0000-0002-2332-8784}, M.~Spanring\cmsorcid{0000-0001-6328-7887}, S.~Templ\cmsorcid{0000-0003-3137-5692}, W.~Waltenberger\cmsorcid{0000-0002-6215-7228}, C.-E.~Wulz\cmsAuthorMark{1}\cmsorcid{0000-0001-9226-5812}
% ...
\vskip\cmsinstskip
% ...
1:~Also at TU Wien, Wien, Austria\\
% ...
\end{sloppypar}
\end{document}